\documentclass[12pt]{article}
\usepackage{eqsection,latexsym,epsf,epsfig,cite,amssymb}

\begin{document}

\date{January 11, 2002}

\title{
25th-order high-temperature expansion results \\
for three-dimensional Ising-like systems \\ 
on the simple cubic lattice.
}
\author{
Massimo Campostrini$\,^1$, Andrea Pelissetto$\,^2$, 
Paolo Rossi$\,^1$, Ettore Vicari$\,^1$ \vspace{2mm}\\
\normalsize\rm
$^1$ Dipartimento di Fisica dell'Universit\`a di Pisa and I.N.F.N., 
I-56126 Pisa, Italy\hfill \\
\normalsize\rm
$^2$ Dipartimento di Fisica dell'Universit\`a di Roma I
and I.N.F.N., I-00185 Roma, Italy\hfill \\
}

\maketitle

\begin{abstract}
25th-order 
high-temperature series are computed for a general
nearest-neigh\-bor three-dimensional Ising
model with arbitrary potential on the simple cubic lattice.  
In particular, we consider three improved potentials 
characterized by suppressed leading scaling corrections.
Critical exponents are extracted from high-temperature
series specialized to improved potentials, obtaining
$\gamma=1.2373(2)$, $\nu=0.63012(16)$, $\alpha=0.1096(5)$, 
$\eta=0.03639(15)$, $\beta=0.32653(10)$, $\delta=4.7893(8)$.  
Moreover, biased analyses of the 25th-order series of the
standard Ising model provide the estimate $\Delta=0.52(3)$
for the exponent associated with the leading scaling corrections.

By the same technique, we study 
the small-magnetization expansion of the Helmholtz free energy.  The
results are  then applied to the construction of parametric representations
of the critical equation of state, using a systematic approach
based on a global stationarity condition. 
Accurate estimates of several universal amplitude  ratios
are also presented.
\end{abstract}



\newpage


\section{Introduction}
\label{intro}

The Ising model is one of the most studied systems in the 
theory of phase transitions,
not only because it is the simplest nontrivial model that has  a critical 
behavior with nonclassical exponents, 
but also because it describes the critical behavior of many physical
systems.  Indeed, many 
systems characterized by short-range interactions and a scalar order
parameter undergo a continuous phase transition belonging 
to the Ising universality class. We mention 
the liquid-vapor transition in simple fluids and the 
critical transitions in multicomponent fluid mixtures, in uniaxial
antiferromagnetic materials, and in micellar systems.
Continuous transitions belonging to the three-dimensional Ising universality
class are also expected in high-energy
physics, for instance in the electroweak theory at finite temperature and
in the theory of strong interactions at 
finite temperature and finite baryon-number chemical potential.   
For a recent review, see, e.g., Ref.~\cite{PV-review}.

The high-temperature (HT) expansion is one of the most efficient
approaches to the study of critical phenomena. Very precise results 
have been obtained by performing careful extrapolations to the 
critical point, by using several different methods, see, e.g., 
Ref. \cite{Guttmann-rev-89}.
For moderately long series, such as those available for models
in the three-dimensional Ising universality class,
the nonanalytic confluent corrections are the 
main source of systematic errors.
For instance, according to renormalization-group theory,
the critical behavior of the
magnetic susceptibility is given by the Wegner expansion
\begin{equation}
\chi = C t^{-\gamma} \left( 1 + a_\chi \,t^\Delta + a_2\, t^{2\Delta} + ...
+ b\, t^{\Delta_2} + ... + e_1\, t + e_2\, t^2 + ... \right),
\label{chiwexp}
\end{equation}
where $t\equiv (T-T_c)/T_c$ is the reduced temperature
and $\Delta$ is a noninteger exponent, $\Delta\approx 0.5$ 
in the Ising case.
In the analysis of HT expansions
these nonanalytic terms introduce large and dangerously
undetectable systematic deviations in the results.

In order to obtain precise estimates of the critical parameters, the
approximants of the HT series should properly allow for the confluent
nonanalytic corrections
\cite{ZJ-79,Nickel-82,Gaunt-82,CFN-82,Adler-83,GR-84,FC-85}.   
However, the extensive numerical work that has been done
shows that in practice, with the series of moderate length that 
are available today, no unbiased analysis is able to 
take effectively into account nonanalytic correction-to-scaling
terms. In order to treat them properly, one should use
biased methods in which the presence of the leading  nonanalytic term 
with exponent $\Delta$ is imposed
(see, e.g.,\ Refs.\ 
\cite{Roskies-81,AMP-82,BC-97,PV-98-g,BC-98,BC-00,MJHMJG-00,BC-02}).
An alternative approach to this problem consists in considering models---we
call them {\em improved}---that do not couple the leading irrelevant operator 
that gives rise to the confluent correction of order $t^\Delta$.
Therefore, such correction 
does not appear in the expansion of {\em any}
thermodynamic quantity near the critical point: for instance,
$a_\chi=0$ in Eq.~(\ref{chiwexp}).
In this case, we expect standard analysis techniques
to be much more effective,
since the main source of systematic error should have been eliminated.
There are no methods that allow to determine exactly improved models, and one
must therefore use numerical techniques. One may use HT expansions,
but in this case the improved model is determined  with a relatively large
error \cite{CFN-82,GR-84,FC-85,NR-90,CPRV-99,PV-review} so that
the final results do
not significantly improve the estimates obtained from standard
analyses using biased approximants.
Recently
\cite{BFMM-98,HPV-99,BFMMPR-99,Hasenbusch-99,CPRV-99,HT-99,CHPRV-01,%
Hasenbusch-01,CHPRV-02},
it has been realized that Monte Carlo (MC) simulations using
finite-size scaling techniques are very effective for this purpose,
obtaining accurate determinations of
several improved models in the Ising, XY, and O(3) universality
classes. 

As shown in Refs.~\cite{CPRV-99,CPRV-00,CPRV-00-2,CHPRV-01,CHPRV-02},
analyses of the HT series for the improved models
lead to a significant improvement in the estimates of the 
critical exponents and of other infinite-volume HT quantities.
Our working hypothesis is that, with the series of current length,
the systematic errors, i.e., the systematic deviations
that are not taken into account in the analysis, are largely due to the
leading confluent correction, so that improved models 
give results with smaller and, more importantly, reliable error estimates. This
hypothesis can be checked by comparing the results obtained using
different improved models: if correct, they should agree within error bars.
In the following we shall report results that confirm
our hypothesis. Indeed,
the estimates obtained from three different improved
Hamiltonians are perfectly consistent.
Moreover, they are very stable with
respect to the order of the series considered in the analysis,
without showing dangerous trends, but only an apparent
reduction of the error.  The results obtained in Ref.~\cite{CPRV-99}
using 20th-order series are fully consistent with
the 25th-order analysis that we present.

We consider a simple cubic lattice and scalar models with Hamiltonian
\begin{equation}
{\cal H} = -\beta \sum_{<i,j>} \phi_i \phi_j + 
\sum_i V(\phi_i^2),\label{hamiltonian}
\end{equation}
where $\beta\equiv 1/T$, ${<}i,j{>}$ indicates nearest-neighbor sites,
$\phi_i$ are real variables, and $V(\phi^2)$ is a generic potential
satisfying appropriate stability constraints.  
These models are expected to have either a critical transition 
belonging to the Ising universality class or a first-order transition
between a disordered and an ordered phase, 
apart from special cases that correspond to multicritical points.
Using the linked-cluster expansion technique, 
we compute, for an arbitrary potential,  the 
HT expansion of the
two-point correlation function to 25th order on a simple cubic lattice.
These results extend those of Ref. \cite{CPRV-99} that reported 
the two-point function to 20th order \cite{bcc-series}.
In particular, we consider 
three classes of models depending on an irrelevant parameter,
which is fixed by requiring the absence of the leading scaling correction.
The first one is the $\phi^4$ lattice model with potential
\begin{equation}
V(\phi^2) = \phi^2 + \lambda_4 (\phi^2-1)^2. 
\label{potentialphi4}
\end{equation}
MC simulations using finite-size scaling techniques
have shown that the model is improved for 
\cite{phi4est}
\begin{equation}
\lambda_4=\lambda_4^*=1.10(2).
\end{equation}
A consistent but less precise estimate can be obtained from the
HT expansion \cite{CPRV-99}.
The second class of models is the $\phi^6$ lattice model
with potential 
\begin{equation}
V(\phi^2) = \phi^2 + \lambda_4 (\phi^2-1)^2 + \lambda_6(\phi^2-1)^3. 
\label{potential}
\end{equation}
Fixing $\lambda_6=1$, the $\phi^6$ Hamiltonian is improved for \cite{CPRV-99}
\begin{equation}
\lambda_4=\lambda_4^*=1.90(4).
\end{equation}
Finally, we consider the spin-1 (or Blume-Capel) Hamiltonian
\begin{equation}
{\cal H} = -\beta \sum_{<i,j>}s_i s_j + D\sum_i s^2_i,
\label{spin1action}
\end{equation}
where the variables $s_i$ take the values $0,\pm 1$.
An improved spin-1 model is obtained for
\cite{Hasenbusch-99-h} 
\begin{equation}
D=D^*=0.641(8).
\end{equation}
The comparison of the results obtained using the 
above-mentioned improved Hamiltonians 
represents a strong check of the expected
reduction of systematic errors in the HT results,
and provides an estimate of the residual errors due to the subleading
confluent corrections to scaling.

We also extend the  HT expansion of the zero-momentum $n$-point
correlation functions  $\chi_n$. In particular, we compute $\chi_4$, $\chi_6$,
and $\chi_8$  to 21st, 19th, and 17th order respectively.
The analysis of such series provides information on the
small-magnetization expansion of the Helmholtz free energy in
the HT phase. These results are used to
determine  approximate representations of the equation of state that 
are valid in the critical regime in the whole $(t,H)$ plane. 
For this purpose, 
following Ref. \cite{CPRV-99}, we use  a systematic approximation 
scheme based on polynomial parametric representations
and on a global stationarity condition.
This approach allows us to obtain an accurate determination of the
critical equation of state in the whole 
critical region  up to the coexistence curve.

In Table~\ref{summary} we anticipate most of the results that we shall obtain
in this paper. We report HT estimates of the critical exponents and of the 
coefficients parametrizing the small-magnetization
expansion of the Helmholtz free energy: they are 
denoted by IHT, where the ``I" stresses the fact that we are using 
improved models. Then, we report several amplitude ratios 
(definitions are given in Sec. \ref{univratio}). Those appearing in the 
column IHT-PR are obtained from
an approximate representation of the equation of
state that uses the HT results as inputs,
those labelled by LT are obtained from the analysis
of low-temperature expansions, while those reported under IHT-PR+LT
are obtained combining the IHT-PR and LT results. 
The comparison with the corresponding  Table XIII of
Ref.~\cite{CPRV-99} shows that the estimates obtained from the 
25th-order series are essentially identical to those obtained by 
using the shorter 20th-order series. However, the longer series allow us 
to give error bars that are smaller by a factor of 1.5-2, depending on the 
observable.
The estimates reported in Table~\ref{summary} are 
in substantial agreement with, and substantially more precise than,
the best theoretical and experimental 
results that have been previously obtained
\cite{BB-85,BBMN-87,LF-89,GKM-96,ZF-96,CH-97,GZ-97,FZ-98,GZ-98,HP-98,LMSD-98,%
BFMM-98,PV-98,HPV-99,BFMMPR-99,Hasenbusch-99,BST-99,PV-99,BC-00,ZJ-01,NGMJ-01}. 
For a comprehensive recent review of theoretical and experimental results,
see Ref.~\cite{PV-review}.
On the experimental side, we mention the planned experiments
in microgravity environment described in Ref. \cite{MISTE}, which  
may substantially improve the experimental determinations of the
critical quantities and make the comparison with the theoretical
computations more stringent.

After completion of this work, the study reported in Ref.~\cite{BC-02}
appeared, where analyses of 25th-order series for spin-$S$ models are
reported.  Results for the critical exponents are obtained by means of
biased analyses, essentially by fixing $\Delta$.  Comparing
Ref.~\cite{BC-02} with Refs.~\cite{BC-97,BC-00}, where 21st- and
23rd-order series are analyzed, a trend appears towards better
agreement with improved Hamiltonian results (Ref.~\cite{CPRV-99} and
present paper).  The latest results of the authors of
Ref.~\cite{BC-02} are in full agreement with our estimates.

\begin{table*}
\footnotesize
\begin{center}
\caption{
Summary of the results obtained in this paper (unless a reference is explicitly
cited) by our high-temperature
calculations (IHT), by using the parametric representation of the
equation of state (IHT-PR), by analyzing the low-temperature expansion
(LT), and by combining the results of the
two approaches (IHT-PR+LT).
The estimates of critical exponents marked by an asterisk have been obtained using
scaling and hyperscaling relations. 
}
\label{summary}
\begin{tabular}{|c|c|cccc|}
\hline\hline
\multicolumn{1}{l}{}&
\multicolumn{1}{l}{}&
\multicolumn{1}{c}{IHT}&
\multicolumn{1}{c}{IHT-PR}&
\multicolumn{1}{c}{LT}&
\multicolumn{1}{c}{IHT-PR+LT}\\
\hline \hline
critical 
&$\gamma$ & 1.2373(2) &  &  & \\
exponents
&$\nu$    & 0.63012(16) &  &  &  \\
&$\alpha$ & 0.110(2),$\;^*$0.1096(5)  &  &  &  \\
&$\eta$   & $^*$0.03639(15)   &  &  & \\
&$\beta$  & $^*$0.32653(10)   &  &  & \\
&$\delta$ & $^*$4.7893(8)   &  &  & \\
&$\Delta$ & 0.52(3)   &  &  & \\
&$\omega$ & 0.83(5)   &  &  & \\
&$\omega_{\rm NR} $ & 2.0208(12) \cite{CPRV-99,foot-omegaNR}  &  &  & \\
\hline

small-magnetization
&$g_4^+$ & 23.56(2) && &  \\
expansion of
&$r_6$    & 2.056(5)    &  &  & \\
the free-energy
&$r_8$    & 2.3(1)     &  &  & \\
in the HT phase
&$r_{10}$ &$-$13(4) \cite{CPRV-99}    & $-$10.6(1.8)  & & \\
\hline

universal 
&$U_0$ & &  0.532(3) & &\\
amplitude
&$U_2$ & &  4.76(2) & & \\
ratios
&$U_4$ &  & $-$9.0(2) & & \\
see Sec. \ref{univratio}
&$R_c^+$ &  & 0.0567(3) & & \\
for notations
&$R_c^-$ &  & 0.02242(12) && \\

&$R_4^+$&  & 7.81(2) & &\\
&$v_3$  &  & 6.050(13) & &\\
&$R_4^-$ &  & 93.6(6) && \\

&$v_4$ &  & 16.17(10) && \\

&$R_\chi$ &  & 1.660(4) && \\

&$w^2$ & &  & 4.75(4) \cite{PV-98-g} &   \\ 

&$U_\xi$ &  && & 1.956(7) \\
&$Q^+ $  &   & 0.01880(8) & & \\

&$R^+_\xi$  &  &  0.2659(4) &&  \\
&$Q^- $  & & & &  0.00472(5) \\
&$Q_c $ &  & 0.3315(10) &&     \\
&$g_3^-$ &   & & & 13.19(6) \\
&$g_4^-$ &  & &   & 76.8(8) \\
&$Q^+_\xi$ & 1.000200(3) && &\\
&$Q^-_\xi$ & && 1.032(4) \cite{CPRV-98} &  \\
&$U_{\xi_{\rm gap}}$ & && & 1.896(10) \\
&$Q^c_\xi$ & && & 1.024(4) \\
&$Q_2$ & && & 1.195(10) \\
&$P_m$ &  & 1.2498(6) & & \\
&$P_c$ &  & 0.3933(7) & & \\
&$R_p$ &  & 1.9665(10) & & \\

\hline\hline
\end{tabular}
\end{center}
\end{table*}

The paper is organized as follows.  

In Sec.\ \ref{HTexp} we report the HT expansions. 
Section \ref{crexpHT} reports the results of our 
analysis of the HT series for the critical exponents.
In Sec.\ \ref{CES} we determine approximate
representations of  the critical equation of state.
In Sec. \ref{CES.1} we give the definitions,
in Sec.\ \ref{IHTrj} 
we give estimates of the
zero-momentum four-point coupling and of the
first few coefficients of the small-magnetization expansion of the 
equation of state, in Sec. \ref{appeq} we explain the method, and 
in Sec. \ref{reseq} we give the final results.
In Sec.~\ref{univratio} we present estimates of
several universal amplitude ratios.
In Sec.\ \ref{twopointf} we determine the low-momentum behavior of
the two-point function in the HT phase.

\section{High-temperature expansion}
\label{HTexp}

We considered a simple cubic lattice and computed the HT expansion of
several quantities for a generic lattice model defined by the
Hamiltonian (\ref{hamiltonian}), using the vertex- and
edge-renormalized linked-cluster expansion technique, developed in
Refs.~\cite{Wortis-74,NR-90} and described in detail in
Ref.~\cite{Campostrini-01}.  Some technical points that allowed us to
extend the computation of Ref.~\cite{Campostrini-01} will be reported
in a forthcoming publication.  We computed the 25th-order HT expansion
of the two-point function
\begin{equation}
G(x) = \langle \phi(0) \phi(x) \rangle.
\end{equation}
In the present context we consider its moments
\begin{equation}
m_{2j} = \sum_x |x|^{2j} \, G(x),
\label{m2j}
\end{equation}
and therefore, the magnetic susceptibility  $\chi\equiv m_0$ and
of the second-moment correlation length $\xi^2 = m_2 /(6\chi)$.

We also calculated the HT expansion of the zero-momentum
connected $2j$-point correlation functions $\chi_{2j}$
\begin{equation}
\chi_{2j} = \sum_{x_2,...,x_{2j}}
    \langle \phi(0) \phi(x_2) ...
        \phi(x_{2j-1}) \phi(x_{2j})\rangle_c
\end{equation}
($\chi = \chi_2$). More precisely, we computed $\chi_4$ to 21st order,
$\chi_6$ to 19th order, $\chi_8$ to 17th order. The correlation function
$\chi_{10}$ was
computed to 15th order in Ref.~\cite{CPRV-99}.

It would be pointless to present here the full results for an
arbitrary potential: the resulting expressions are only fit for
further computer manipulation. They are available 
on request.
In Table \ref{series} we give the new coefficients only for 
the three improved models we have
considered, i.e., for the $\phi^4$ model at $\lambda_4=1.10$,
for the $\phi^6$ model at $\lambda_6=1$ and $\lambda_4=1.90$,
and for the spin-1 model at $D=0.641$.

For the standard Ising model, 
we give below the coefficients of the terms that extend the expansions 
presented in Refs.~\cite{Campostrini-01,BC-98} for $\chi$, $m_2$, and $\chi_4$:
\begin{eqnarray}
\chi &=& \cdots + 18554916271112254v^{24} +  85923704942057238 v^{25} + 
        O(v^{26}) \nonumber \\
m_2 &=& \cdots +  977496788431483776 v^{24} + 4767378698515169334 v^{25} +
        O(v^{26}), \nonumber \\
\chi_4 &=& \cdots -6306916133817628 v^{18}
    -34120335459595728 v^{19} \nonumber \\
       && -183166058308506108 v^{20}  -976373577976196368 v^{21} +  O(v^{22}),
\end{eqnarray}
where $v\equiv \tanh\beta$.

\begin{table}
\footnotesize
\hspace*{-1.5cm}    
\caption{Coefficients of the HT series for the improved models. 
Lower-order coefficients appear in Ref. \cite{CPRV-99}.
 }
\label{series}
\begin{tabular}{lccc}
\hline\hline
$n$ & $\phi^4:\quad \lambda_4=1.10$ 
& $\phi^6:\quad \lambda_6=1, \;\lambda_4=1.90$ & 
spin-1$:\quad D=0.641$ \\
\hline
\multicolumn{4}{c}{$\chi$} \\
21 & 958465949.119795229380125  &  55356759.0258594943774739  &  
     521863527.549747127784405 \\ 
22 & 2581828793.17418316658592  &  130996257.131383657648562  &  
     1367254366.70256684609648 \\ 
23 & 6953921835.10625772660286  &  309956395.981892002096689  &  
     3581814299.63029965928082 \\ 
24 & 18716342130.2600278822297  &  732873665.558914443007657  &  
     9376338630.49601545283933 \\ 
25 & 50369768053.5367726030130  &  1732674465.68758001711514  &  
     24543094928.9205155990856 \\ 
\hline
\multicolumn{4}{c}{$m_2$} \\
21 & 32990320251.5660972216018  &  1900950559.23375555678011  &  
     17908950773.4801706544197 \\ 
22 & 94071328367.8146359923071  &  4762044317.91673448231502  &  
     49684326561.5439542757331 \\ 
23 & 267461898855.689392585599  &  11894571003.1970044574018  &  
     137433163639.457494472451 \\ 
24 & 758423675496.642760823002  &  29631147101.2512233682029  &  
     379139772127.101469600055 \\ 
25 & 2145329356955.42924803892  &  73634162230.2093808561076  &  
     1043350926215.22611634874 \\ 
\hline
\multicolumn{4}{c}{$m_4$} \\
20 & 541141652908.631074719231  &  35399348720.3598637148375  &  
     299758906549.791610350073 \\ 
21 & 1643345014677.80358819408  &  94444621918.7858920241050  &  
     885976701269.736104292700 \\ 
22 & 4961021084766.33884428748  &  250485298262.046958470064  &  
     2603026564263.78069815384 \\ 
23 & 14895796670810.3387628037  &  660748522303.208118944668  &  
     7606210964865.32821158574 \\ 
24 & 44504475774409.2126174407  &  1734347627024.93369651634  &  
     22115153167519.1984380502 \\ 
25 & 132362288688779.709839376  &  4531641133142.45499870752  &  
     64005596692608.8036008995 \\ 
\hline
\multicolumn{4}{c}{$\chi_4$} \\
19 & $-$141558376231.985023846408  &  $-$9210343000.40488445467068  &  
     $-$77210883309.3840433243811 \\ 
20 & $-$440895445559.088001425635  &  $-$25206881115.0765162521666  &  
     $-$234263398532.544236218037 \\ 
21 & $-$1363771989486.31756523825  &  $-$68511054288.5805997438372  &  
     $-$705801443484.646787710146 \\ 
\hline
\multicolumn{4}{c}{$\chi_6$} \\
18 & 25922773662329.4681285982  &  1657400403425.39611029038  &  
     13110582140461.8241625980 \\ 
19 & 93214547843378.1420243052  &  5239283130720.37310719268  &  
     46080008679021.7095625364 \\ 
\hline
\multicolumn{4}{c}{$\chi_8$} \\
17 & $-$3021378127745877.943411840  &  $-$188904527250502.5683919596  &  
     $-$1360671334948122.792253527 \\ 
\hline\hline
\end{tabular}
\end{table}

\section{The critical exponents}
\label{crexpHT}

In this section  we shall report three different analyses for the 
critical exponents. In Sec. \ref{iaan} we shall use integral 
approximants and derive the estimates reported in Table \ref{summary}.
In Secs. \ref{rmet} and \ref{matching} we shall use two other methods 
that have been recently used in the literature
\cite{MJHMJG-00,BC-00,BC-02} to confirm the integral-approximant results.

\subsection{Analysis using integral approximants}
\label{iaan}

In order to estimate $\gamma$ and $\nu$, we analyze
the HT series of the magnetic susceptibility and of the
second-moment correlation length  respectively. 
We follow closely App. B of Ref. \cite{CPRV-99},
to which the reader is referred for more details.

We  use  integral approximants (IA's) of
first, second, and third order (see Ref.~\cite{Guttmann-rev-89} for a review).  
Given an $n$th-order series $f(\beta)= \sum_{i=0}^n c_i \beta^i$, its
$k$th-order integral approximant $[m_k/m_{k-1}/.../m_0/l]$ IA$k$ is a
solution of the inhomogeneous $k$th-order linear differential equation
\begin{equation}
P_k(\beta) f^{(k)}(\beta) + P_{k-1}(\beta) f^{(k-1)}(\beta) + ... 
+ P_1(\beta)f^{(1)}(\beta)+ P_0(\beta)f(\beta)+R(\beta)= 0,
\label{IAkdef}
\end{equation}
where the functions $P_i(\beta)$ and $R(\beta)$ are polynomials of
order $m_i$ and $l$ respectively, which are determined by the known
$n$th-order small-$\beta$ expansion of $f(\beta)$.
Following Fisher and Chen \cite{FC-85}, we also consider 
integral approximants, FCIA$k$'s, in which 
$P_k(\beta)$ is a polynomial in $\beta^2$.  
FCIA$k$'s  allow for the presence of the antiferromagnetic singularity 
at $\beta_c^{\rm af}=-\beta_c$ \cite{Fisher-62}.
In our analyses we consider diagonal or quasi-diagonal approximants,
since they are expected to give the most accurate results.
For each set of IA$k$'s we determine  the average of the values
corresponding to all nondefective IA$k$'s. 
The error bar from each class of IA's is essentially 
the spread of the results, and it is given by the standard deviation 
of the results obtained from all nondefective IA's.
In most cases the nondefective IA's are more than 90\%.

\begin{figure}[tbp]
\hspace{-1cm}
\vspace{0cm}
\centerline{\psfig{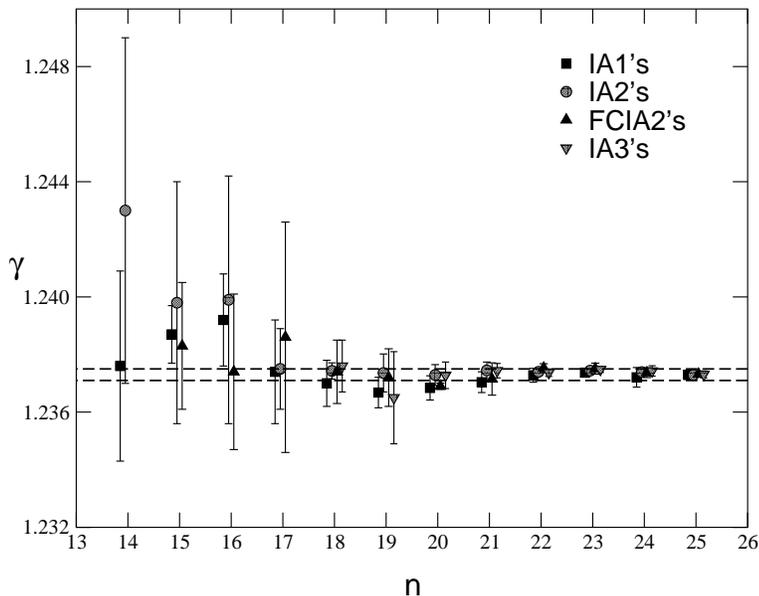}}
\vspace{0cm}
\caption{
Estimates of $\gamma$ as obtained
by analyzing the HT series of $\chi$ for the $\phi^4$ model at 
$\lambda_4=1.10$ versus the 
order $n$ of the series considered in the analysis.
Several approximants (defined in the text) are considered.
}
\label{phi4IAs}
\end{figure}

\begin{figure}[tbp]
\hspace{-1cm}
\vspace{0cm}
\centerline{\psfig{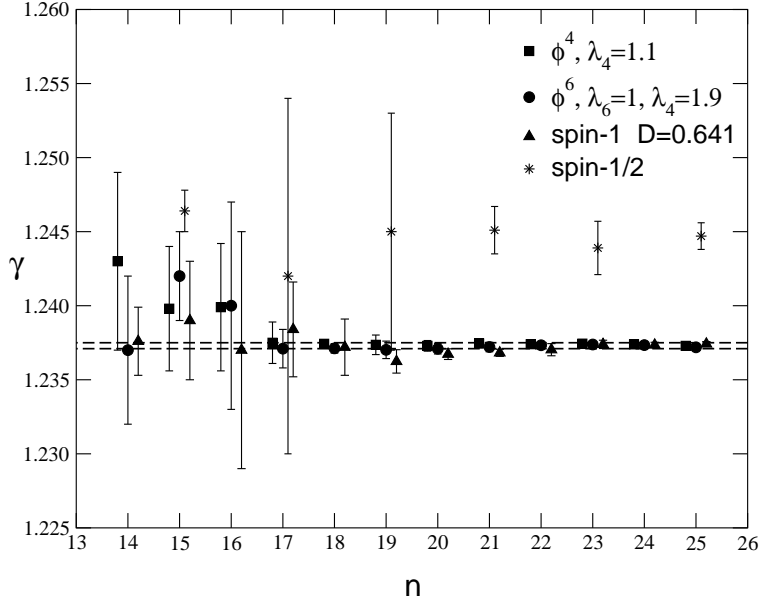}}
\vspace{0cm}
\caption{
Estimates of $\gamma$ as obtained
by analyzing the HT series of $\chi$ for the improved models and for the 
spin-1/2 model,
versus the order $n$ of the series considered in the analysis.  
IA2's are considered.
}
\label{IA2s}
\end{figure}

All IA's considered give perfectly consistent results. 
Moreover, the results turn out to be very  stable with respect to the
number of terms of the series, so that there is no need to perform
problematic extrapolations in the number of terms in order to obtain
the final estimates.  
In Fig.~\ref{phi4IAs} we show the estimates of $\gamma$ obtained
by analyzing the series of $\chi$ for the $\phi^4$ model at $\lambda_4=1.10$
by using 
IA1's, IA2's, IA3's, and FCIA2's, as a function of the 
order $n$ of the series considered in the analysis. 
Perfect agreement is also found among the results for the three improved
Hamiltonians. This is shown in Fig.~\ref{IA2s}, where the results of
the IA2 analyses for the three improved Hamiltonians are reported
versus $n$. 
In Fig.~\ref{IA2s} we also show the results of the IA2 analysis
applied to the series of $\chi$ for the standard Ising spin-1/2 model.
The corresponding results disagree with those obtained by using 
improved Hamiltonians: clearly, there is a large error
that is not taken into account by the spread of the approximants.
The results for the Ising model improve if one biases the analysis
by using the very accurate MC estimate
of $\beta_c$ \cite{BST-99}:  $\beta_c=0.22165459(10)$. Indeed,
$\gamma$ drops from 1.245 to $\gamma=1.2400(5)$. 
However, the error obtained from the spread of the approximants is still
incorrect. Results that are closer to those obtained by using the 
improved Hamiltonians (and substantially compatible with them)
are only obtained by additionally biasing the series, 
allowing for $O(t^\Delta)$ confluent corrections,
see, e.g., Ref.~\cite{BC-00}.

\begin{table}[tbp]
\footnotesize
\begin{center}
\caption{
Critical exponents obtained from
the HT analysis. In parentheses we report the approximant error 
at $\lambda^*$ or $D^*$, in brackets the uncertainty due to the 
error on $\lambda^*$ or $D^*$, in braces the uncertainty due to the
error on $\beta_c$.
}
\label{critexponents}
\begin{tabular}{ccccc}
\hline\hline
\multicolumn{1}{l}{}&
\multicolumn{1}{l}{}&
\multicolumn{1}{c}{$\phi^4$}&
\multicolumn{1}{c}{$\phi^6$}&
\multicolumn{1}{c}{spin-1}\\
\hline \hline
$\gamma$ & $\chi$-series 
& 1.23732(10)[12] & 1.23726(10)[22] & 1.23725(20)[10] \\
$\alpha$ & $\chi$-series (af) 
& 0.110(2) & 0.110(2)  & 0.112(5)  \\
$\nu$ & $\xi^2$-series 
& 0.6302(2)[1] & 0.6301(3)[3]  & 0.6300(2)[1] \\
$\nu$ & $\xi^2$-series ($\beta_c$-biased) 
& 0.63014(1)\{6\}[9] & 0.63009(1)\{16\}[16]  & 0.63010(1)\{10\}[9] \\
$\eta\nu$ & $\chi,\xi^2$-series (CPRM) 
&0.02294(3)[6] & 0.02291(2)[10] &0.02294(8)[4] \\
\hline\hline
\end{tabular}
\end{center}
\end{table}

For the $\phi^4$ lattice model we obtained
\begin{eqnarray}
&&\beta_c(\lambda_4=1.10) = 0.3750975(5), 
\label{betacphi4}\\
&&\gamma_e(\lambda_4) = 1.23732(10) + 0.006(\lambda_4 - 1.10),
\end{eqnarray}
where $\gamma_e(\lambda_4)$ is the effective critical exponent
obtained in the IA analysis, which has a small but nonvanishing
dependence on $\lambda_4$ around the favorite value $\lambda_4=1.10$.
(Here and in the following, we report explicitly the dependance on
$\lambda_4$ and equivalent couplings: should a better estimate of 
$\lambda_4^*$ become available, it can be immediately used to improve
our results.)
The number between parentheses is basically the spread of
the approximants at $\lambda_4=1.10$. 
The $\lambda_4$-dependence is estimated by
determining the variation of the results
when changing $\lambda_4$ around $\lambda_4=1.10$.
The best estimate of $\gamma$ should be obtained at $\lambda_4=\lambda_4^*$.
Thus, using the MC estimate of $\lambda_4^*$,
i.e., $\lambda_4^*=1.10(2)$, and taking into account
its uncertainty, we obtain the estimate $\gamma=1.23732(10)[12]$ 
(which is also reported in Table~\ref{critexponents}),
where the error in brackets is related to the uncertainty
on $\lambda_4^*$.  As final error we consider, prudentially,
the sum of these two numbers.
The estimate (\ref{betacphi4}) is in substantial agreement with the 
MC estimate of $\beta_c$ \cite{Hasenbusch-99} obtained
using finite-size scaling techniques,
$\beta_c(\lambda_4=1.10)=0.3750966(4)$.

Similarly, 
for the $\phi^6$ lattice model
we obtain
\begin{eqnarray}
&&\beta_c(\lambda_4=1.90,\lambda_6=1) = 0.4269791(5) ,
\label{betac-phi6} \\
&&\gamma_e(\lambda_4,\lambda_6=1) = 1.23726(10) + 0.0055(\lambda_4 - 1.90),
\end{eqnarray}
and, using the MC result $\lambda_4^*=1.90(4)$, the
estimate $\gamma=1.23726(10)[22]$; 
for the spin-1 model
\begin{eqnarray}
&&\beta_c(D=0.641) = 0.3856717(10),
\label{betac-spin-1} \\
&&\gamma_e(D) = 1.23725(20) - 0.012(D - 0.641)
\end{eqnarray}
and therefore,  using $D^*=0.641(8)$,
$\gamma=1.23725(20)[10]$.

Our final estimate of $\gamma$ is obtained by 
combining the results of the three improved Hamiltonians: 
as estimate we take the weighted average of the three results, 
and as estimate of the uncertainty the smallest of
the three errors. According to this rather subjective but reasonable
procedure we obtain
\begin{equation}
\gamma=1.2373(2).
\end{equation}
A direct estimate of the specific-heat exponent $\alpha$
is obtained from the singular behavior of $\chi$ at the antiferromagnetic
critical point $\beta_c^{{\rm af}}=-\beta_c$, since \cite{Fisher-62}
\begin{equation}
\chi = c_0 + c_1 \left( \beta - \beta_c^{\rm af}\right)^{\theta_{\rm af}} + ... 
\label{fisheraf}
\end{equation}
where 
\begin{equation}
\theta_{\rm af}=1-\alpha.
\end{equation}
FCIAk's provide rather precise estimates of $\theta_{\rm af}$. The
corresponding results for $\alpha$ are
reported in the second line of Table~\ref{critexponents}.
No error in brackets is reported 
since the dependence on $\lambda_4,D$ is negligible.
As final estimate we give
\begin{equation}
\alpha=0.110(2).
\label{alphae}
\end{equation}

\noindent
The exponent $\nu$ is obtained from the series of the second-moment
correlation length $\xi$, since $\xi^2 \sim (\beta_c-\beta)^{-2\nu}$.
Unbiased analyses of the 24th-order series of $\xi^2/\beta$ 
provide the results reported in the third line of Table~\ref{critexponents}.
The corresponding estimates of $\beta_c$ are
consistent with those derived from $\chi$, although less precise.
For instance,
for the $\phi^4$ model at $\lambda_4=1.10$  we found $\beta_c=0.375098(2)$.

In order to get a more precise estimate of $\nu$, we follow the
procedure suggested in Ref.\ \cite{Guttmann-rev-89}, i.e., we use the
estimate of $\beta_c$ obtained from $\chi$ to bias the analysis of
$\xi^2$.  For this purpose we use IA's that have a singularity at a
fixed value of $\beta_c$, or, in order to take into account the
antiferromagnetic singularity, a pair of singularities at
$\pm\beta_c$; the two choices give equivalent results.  This analysis
provides the following effective exponents for the three classes of
models. For $\lambda_4\approx \lambda_4^*$
\begin{equation}
\nu_e(\lambda_4) = 0.63014(1)\{6\} + 0.0045 (\lambda_4 - 1.10)
\end{equation}
for the $\phi^4$ model,
where the number in braces gives the variation of the estimate when 
$\beta_c$ varies within one error bar; 
\begin{equation}
\nu_e(\lambda_4,\lambda_6=1) = 0.63009(1)\{16\} + 0.004 (\lambda_4 - 1.90)
\end{equation}
for the $\phi^6$ model;
\begin{equation}
\nu_e(D) = 0.63010(1)\{10\} - 0.011 (D - 0.641)
\end{equation}
for the spin-1 model. Then, using the MC estimates of $\lambda_4^*,
D^*$, one obtains the results reported in Table~\ref{critexponents},
where the error due to the uncertainty on $\lambda_4^*$ and $D^*$ is reported
between brackets. 
They are perfect consistent with the results of the unbiased
analysis, but more precise. 
Combining the results of Table \ref{critexponents} 
as we did for $\gamma$, we
obtain
\begin{equation}
\nu = 0.63012(16).
\end{equation}
Using the hyperscaling relation $\alpha=2-3\nu$, we derive
\begin{equation}
\alpha = 0.1096(5),
\end{equation}
which is fully consistent with, but more precise than,
the direct estimate (\ref{alphae}).

Using the above-reported results for $\gamma$ and $\nu$ and 
the scaling relation  $\gamma=(2-\eta)\nu$,
we obtain  $\eta=0.0364(6)$,
where the error is estimated by considering the errors on $\gamma$ and
$\nu$ as independent, which is of course not true.  We can obtain an
estimate of $\eta$ with a smaller, yet reliable, error by applying the
so-called critical-point renormalization method  \cite{CPRM} 
to the series of $\chi$ and $\xi^2$.
This method provides an estimate for the combination $\eta\nu$.
Proceeding as before, we obtain
\begin{equation}
[\eta\nu]_e(\lambda_4) = 0.02294(3) + 0.003 (\lambda_4-1.10)
\end{equation}
for the $\phi^4$ model,
\begin{equation}
[\eta\nu]_e(\lambda_4,\lambda_6=1) = 0.02291(2) + 0.0025 (\lambda_4-1.90)
\end{equation}
for the $\phi^6$ model, and
\begin{equation}
[\eta\nu]_e(D) = 0.02294(8) - 0.005 (D-0.641)
\end{equation}
for the spin-1 model.
We then obtain the results reported in Table \ref{critexponents},
which lead to an estimate of $\eta$  with a considerably
smaller error: 
\begin{equation}
\eta=0.03639(15).
\end{equation}
Then, by using the scaling relations we obtain
\begin{eqnarray}
\delta &=& {5- \eta\over 1 +\eta}= 4.7893(8), \label{deltaex}\\
\beta &=& {\nu\over 2} \left( 1 + \eta\right) = 0.32653(10), \label{betaex}
\end{eqnarray}
where the error on $\beta$ has been estimated by considering the errors of
$\nu$ and $\eta$ as independent.

Finally, we estimate the exponent $\Delta$. For this purpose,
we analyze the HT expansion of $t^\gamma \chi$ that behaves like
\begin{equation}
t^\gamma \chi = C^+\left( 1  + a_\chi  t^\Delta + ...\right),
\end{equation}
for $t\equiv 1 - \beta/\beta_c\to 0$. We consider the 
spin-1/2 model---here improved models are not useful since 
$a_\chi\approx 0$---fix the exponent $\gamma$ 
to our best estimate, $\gamma = 1.2373$, and use biased IA's that 
are singular at $\beta_c = 0.22165459(10)$, which is the most 
precise MC estimate of the critical point \cite{BST-99}.
We obtain
\begin{equation}
\Delta=0.52(3), \label{deltaest}
\end{equation}
where the error takes into account the uncertainty
on $\beta_c$ and $\gamma$. 
Correspondingly, we obtain $\omega=\Delta/\nu=0.83(5)$.
Consistent results are obtained from the analysis
of the series of $t^{2\nu} \xi^2$, fixing $\nu$ and $\beta_c$.

\subsection{The ratio method}
\label{rmet}

In order to check the above-reported results,
we consider the ratio method 
proposed by Zinn-Justin in  Ref.~\cite{ZJ-79}
(see also Ref.~\cite{Guttmann-rev-89}). Such a method 
has been recently employed 
in Refs.~\cite{BC-00,BC-02}  to analyze the 25th-order HT expansions
of spin-$S$ models on the simple cubic and on 
the body-centered cubic lattice.

According to this method, given a quantity
\begin{equation}
S=\sum_n c_n \beta^n\approx A_S (\beta_c-\beta)^{-\zeta} 
\left[1 + a_S (\beta_c-\beta)^\epsilon + ...\right],
\end{equation}
one considers the sequences
\begin{eqnarray}
\beta_c^{(n)} &=& 
\left( {c_{n-2} c_{n-3} \over c_n c_{n-1}} \right)^{1/4} 
\exp \left[ {s_n + s_{n-2} \over 2 s_n (s_n-s_{n-2}) }\right],
\\
\zeta^{(n)} &=& 1 + 2{s_n + s_{n-2}\over (s_n - s_{n-2})^2 },
\end{eqnarray}
where 
\begin{equation}
s_n = - {1\over 2} \left[  {1\over {\rm ln} (c_n c_{n-4}/c_{n-2}^2)} 
+ {1\over {\rm ln} (c_{n-1} c_{n-5}/c_{n-3}^2)} \right].
\end{equation}
Asymptotically, the two sequences $\beta_c^{(n)}$ and $\zeta^{(n)}$
approach $\beta_c$ and $\zeta$, with corrections
of $O(1/n^{1+\epsilon})$ and $O(1/n^\epsilon)$ respectively.
More precisely, if
\begin{equation}
c_n \approx \beta_c^{-n} n^{\zeta-1} 
\left(A_0 + A_\epsilon n^{-\epsilon}\right)
\end{equation}
for $n\to\infty$, then
\begin{eqnarray}
\beta_c^{(n)} &\approx& \beta_c \left[ 1 + {A_\epsilon\over 2 A_0} 
       \epsilon^2 (\epsilon-1) {1\over n^{1+\epsilon}} \right]
\\
\zeta^{(n)} &\approx& \zeta\left[ 1 + {A_\epsilon\over \zeta A_0}
       \epsilon (\epsilon^2-1) {1\over n^{\epsilon}} \right].
\end{eqnarray}
Note that, if only analytic corrections are present, i.e., $\epsilon=1$,
the convergence is faster
with corrections of order $n^{-3}$ and $n^{-2}$ for $\beta_c$ and $\zeta$:
\begin{eqnarray} 
\beta_c^{(n)} &\approx& \beta_c \left[ 1 - 
   \left( {A_1^2 \over A_0^2} {\zeta-2\over\zeta-1} + {7\over12}(\zeta-1)
   \right) {1\over n^3} \right],
\\
\zeta^{(n)} &\approx& \zeta \left[ 1 - 
   \left( 3 {A_1^2 \over A_0^2} {\zeta-2\over\zeta-1} + {3\over4}(\zeta-1)
   \right) {1\over \zeta n^2} \right].
\end{eqnarray}
\begin{figure}[tbp]
\begin{tabular}{c} 
\hspace{0cm}
\psfig{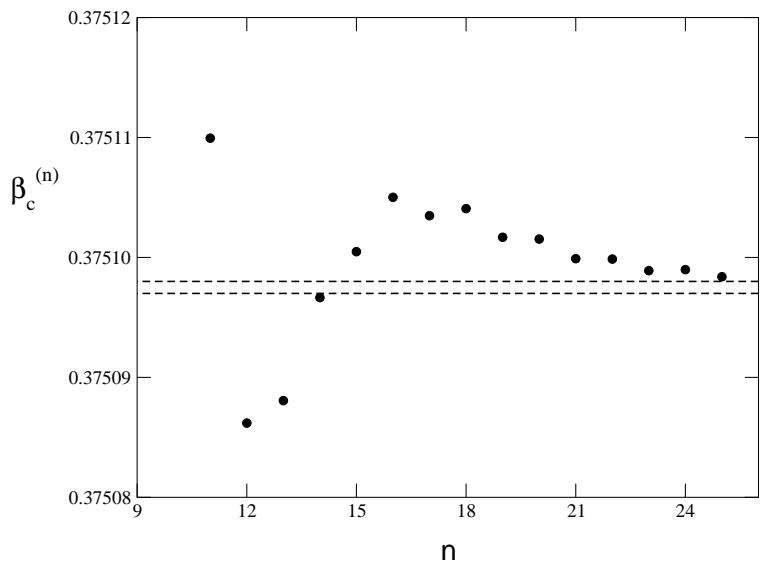} 
\end{tabular}
\caption{
Sequence $\beta_c^{(n)}$ for the
$\phi^4$ model at $\lambda_4=1.10$ using the series for $\chi$.
The dashed lines indicate the IA estimate of $\beta_c$.
}
\label{bcphi4}
\end{figure}
\begin{figure}[tbp]
\begin{tabular}{c} 
\hspace{0cm}
\psfig{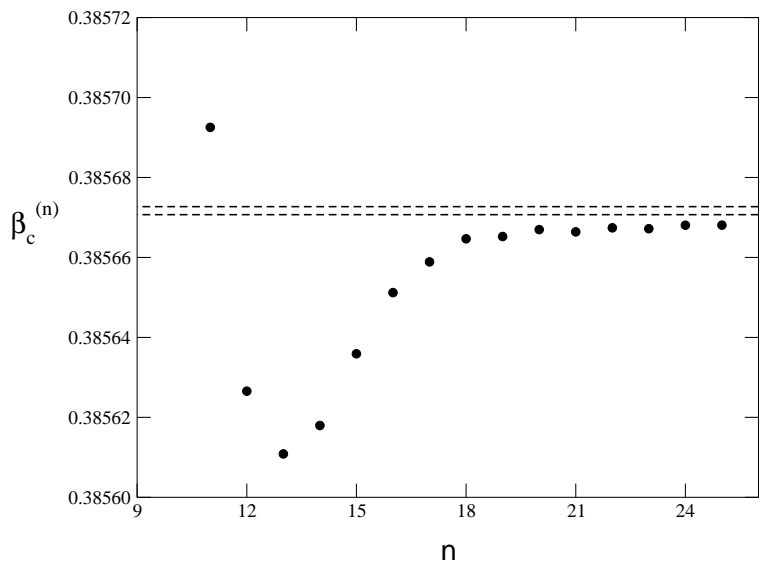} 
\end{tabular}
\caption{
Sequence $\beta_c^{(n)}$ for the
spin-1 model at $D=0.641$ using the series for $\chi$.
The dashed lines indicate the IA estimate of $\beta_c$.
}
\label{bcs1}
\end{figure}
\begin{figure}[tb]
\hspace{-1cm}
\vspace{0cm}
\centerline{\psfig{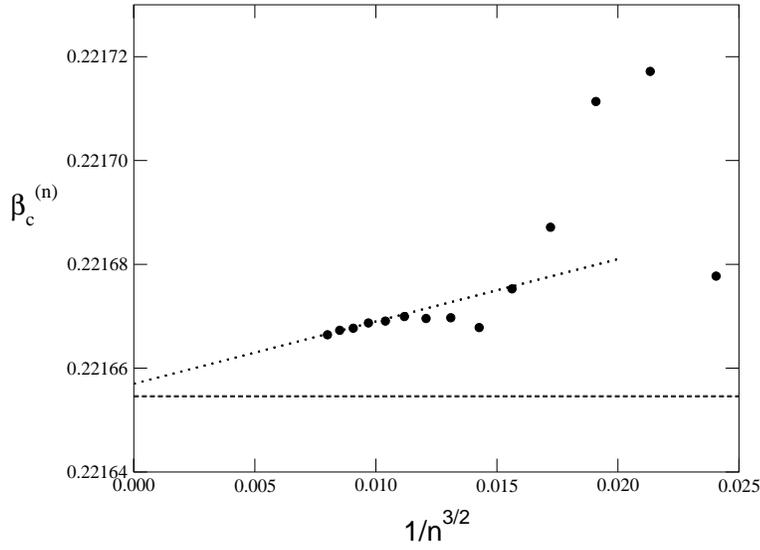}}
\vspace{0cm}
\caption{
Sequence $\beta_c^{(n)}$ for the
spin-1/2 model using the series for $\chi$. The dashed lines indicate
the MC estimate of $\beta_c$, while the dotted line corresponds to 
a $n^{-3/2}$ extrapolation of the four points with $n=22,23,24,25$.
}
\label{bcsisi}
\end{figure}
\begin{figure}[!b]
\hspace{-1cm}
\vspace{0cm}
\centerline{\psfig{width=10truecm,angle=0,file=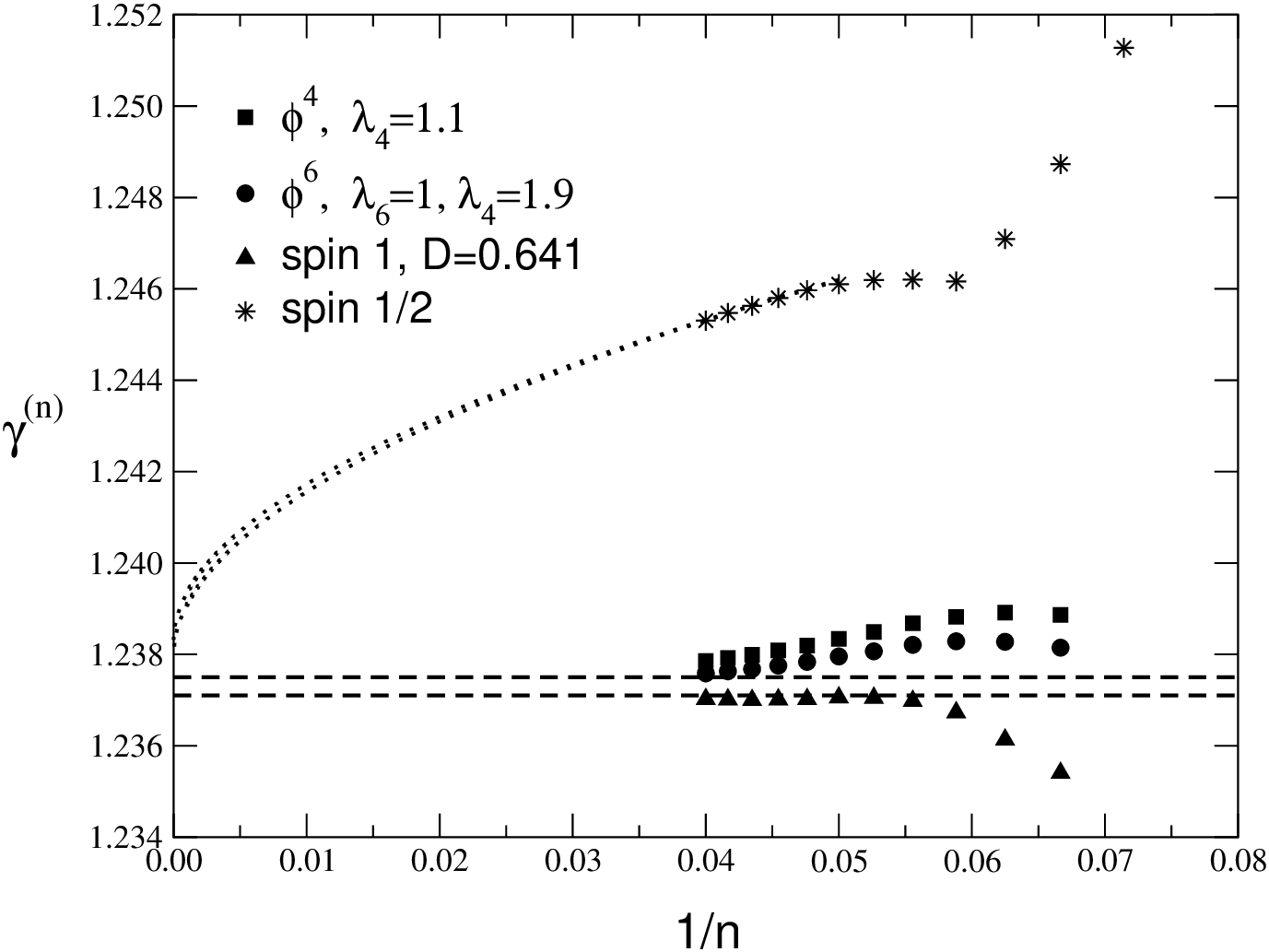}}
\vspace{0cm}
\caption{
Sequences $\gamma^{(n)}$ for 
the $\phi^4$ model at $\lambda_4=1.10$,
the $\phi^6$ model at $\lambda_4=1.90$, 
the spin-1 model at $D=0.641$, and the standard Ising model.
The dashed lines indicate the IA estimate of $\gamma$.
}
\label{gammaratio}
\end{figure}
\begin{figure}[!t]
\hspace{-1cm}
\vspace{0cm}
\centerline{\psfig{width=10truecm,angle=0,file=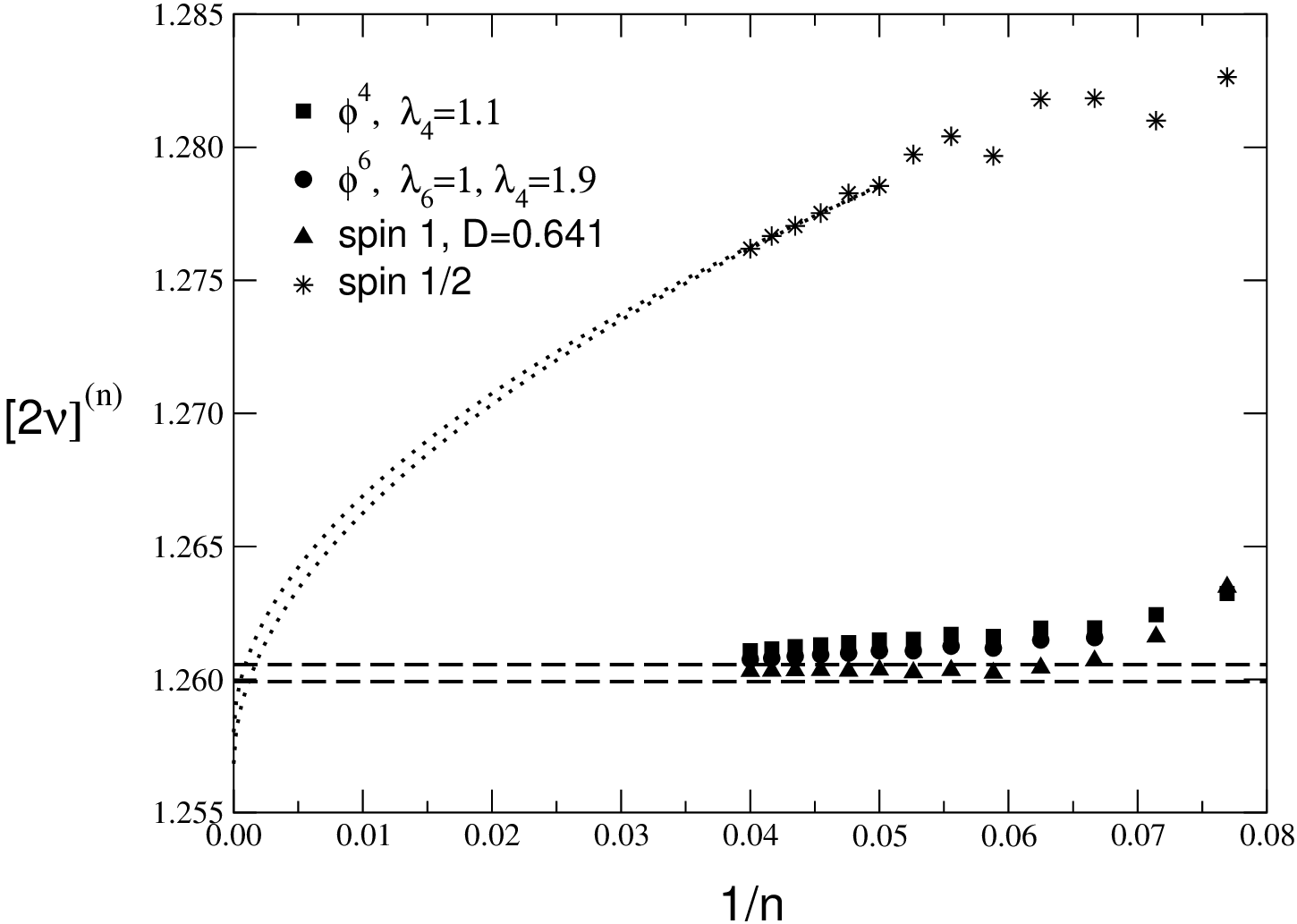}}
\vspace{0cm}
\caption{
Sequences $[2\nu]^{(n)}$ for 
the $\phi^4$ model at $\lambda_4=1.10$,
the $\phi^6$ model at $\lambda_4=1.90$, 
the spin-1 model at $D=0.641$, and the standard Ising model.
The dashed lines indicate the IA estimate of $2 \nu$.
}
\label{nuratio}
\end{figure}
In Figs.~\ref{bcphi4} and \ref{bcs1} we show, for the $\phi^4$ model at
$\lambda_4=1.10$ and for the spin-1 model at $D=0.641$ respectively,
the sequence $\beta_c^{(n)}$ obtained using $S=\chi$.
The sequence clearly approaches the IA estimate.
For the $\phi^4$ model
the agreement is quite good and indeed $\beta_c^{(n)}$ differs from 
the IA estimate (\ref{betacphi4})
 by $15\times 10^{-7}$ and $9\times 10^{-7}$
for $n=24,25$ (note that the error on the IA estimate of 
$\beta_c$ is $5\times 10^{-7}$). 
In principle, one could try to extrapolate the sequence $\beta_c^{(n)}$ 
to get a better estimate of $\beta_c$. For this purpose, we have tried to fit 
$\beta_c^{(n)}$ assuming a behavior of the form
\begin{equation}
\beta_c^{(n)} = a + b n^{-\sigma},
\label{fit-ZJmethod}
\end{equation}
where $a$, $b$, and $\sigma$ are free parameters.
If we interpolate $\beta_c^{(n)}$ for $n=21,23,25$ with 
Eq.~(\ref{fit-ZJmethod}), we obtain
\begin{equation}
\beta_c^{(n)} = 0.3750977(-5) + 3.0(+4)\times 10^{-6} \, 
   \left( {n\over 20}\right)^{-6.6(-1.5)},
\end{equation}
where the ``errors" show the variation of the parameters between the 
interpolation with $n=21,23,25$ and $n=19,21,23$. Analogously, the 
even sequence $n=20,22,24$ gives
\begin{equation}
\beta_c^{(n)} = 0.3750968(-25) + 4(+2)\times 10^{-6} \, 
   \left( {n\over 20}\right)^{-6(-2)}. 
\end{equation}
The extrapolated values are in perfect agreement 
with Eq.~(\ref{betacphi4}), but it is quite difficult to interpret 
the results for $\sigma$. Indeed, in an improved model the leading corrections
in the coefficients $c_n$ are of order $n^{-\Delta_2}$, $n^{-1}$, with 
\cite{NR-84}
$\Delta_2 \approx 1$. The analytic term gives a contribution 
of order $n^{-3}$, while the nonanalytic one gives a correction of 
order $n^{-\Delta_2-1}$. However, its amplitude is of order 
$\Delta_2 - 1$, and thus, since $\Delta_2 \approx 1$, it could be very small.
The next correction terms are of order $n^{-\Delta_3}$, 
$n^{-1-\Delta}$, and give 
rise to corrections of order $n^{-1-\Delta_3}$, 
$n^{-2-\Delta}$. Inclusion of corrections with $2<\sigma \lesssim 5/2$
does not improve the fit \cite{foot-correction}.
Clearly, we are not yet sufficiently asymptotic to be 
able to extrapolate using the leading asymptotic behavior. 
At the values of $n$ we are considering, several 
corrections are still important and apparently conspire 
to give a uniformly small correction. 

The same behavior is observed in the $\phi^6$ model, where both 
odd and even points
extrapolate to 0.4269787, with 
effective exponent $\sigma \approx 12$, 8. The agreement with the 
IA estimate (\ref{betac-phi6})
is quite good. We finally analyze the spin-1 results. 
Even points show again a very fast convergence with $\sigma\approx9$ and 
extrapolate to 0.3856662. Odd points instead are well fitted by assuming 
corrections of order $n^{-2}$ or $n^{-5/2}$. Fixing 
$\sigma=-2$, we obtain 0.3856730, while for $\sigma=-5/2$ we 
have 0.3856719. Again, the IA result (\ref{betac-spin-1}) 
is very well confirmed.

For comparison, in Fig.~\ref{bcsisi} we plot the
sequence $\beta_c^{(n)}$ for the spin-1/2 model versus $1/n^{3/2}$
which should be approximately the leading correction.
The higher-$n$ results have apparently the predicted $O(n^{-3/2})$ behavior,
and indeed an extrapolation with Eq. (\ref{fit-ZJmethod}) and $\sigma=3/2$
gives results that are close to the MC estimate of $\beta_c$. 
The odd (resp. even) points extrapolate to 0.22165686 (resp. 
0.22165717): they are close to the MC estimate  \cite{BST-99}
0.22165459(10).  However, it is hard to 
go beyond a relative precision of $10^{-5}$.

In Fig.~\ref{gammaratio} we show the sequence $\gamma^{(n)}$
as obtained from the series of $\chi$ for the three improved
models and for the standard Ising model.
The improved results clearly approach our best estimate $\gamma=1.2373(2)$,
the $\phi^4$ and $\phi^6$ models from above
and the spin-1 model from below. Note that the results are extremely 
flat and no extrapolation is needed.
We also report the sequence $\gamma^{(n)}$ for the Ising model. 
If we extrapolate the results assuming a behavior of the form 
$a + b n^{-\Delta}$, with $\Delta = 0.52$, we obtain 
$\gamma = 1.23857$, 1.23832, 1.23801 using pairs 
$n=(21,23)$, $(22,24)$, and $(23,25)$. Clearly, the estimates converge 
towards the IA estimate $\gamma = 1.2373(2)$. 

In Fig.~\ref{nuratio}
we show the sequence $[2\nu]^{(n)}$ obtained from the series of $\xi^2$.
Again, the improved models show a very good convergence to the IA 
estimate, in spite of the fact that the analysis is unbiased---the value
of $\beta_c$ is not fixed.
The Ising results are sensibly higher and steadily decreasing, 
reaching $\nu \approx 0.638$ for 
$n = 25$. Results that are closer to the IA estimate are 
obtained by an extrapolation. Assuming a behavior of the form
$a + b n^{-\Delta}$, we obtain $\nu = 0.6290$ and 0.6284
from even and odd sequences respectively. Again, the agreement is 
satisfactory.

In conclusion, this analysis based on the variant 
of the ratio method proposed by Zinn-Justin \cite{ZJ-79}
supports the IA estimate obtained in Sec.~\ref{iaan}.

\subsection{Matching the coefficients with their asymptotic form}
\label{matching} 

In the preceding section we have determined the critical exponents 
and $\beta_c$ by generating sequences that converge to the asymptotic 
value for $n\to \infty$. In this Section, following Ref.~\cite{MJHMJG-00},
we wish to perform a more straightforward analysis, both conceptually and 
practically. The idea is to generate sequences of estimates
by fitting the expansion coefficients with their asymptotic form. 
By adding a sufficiently large number of terms we can make the 
convergence as fast as possible, although of course the procedure becomes
unstable if the number of terms included is too large compared to the 
number of available terms. In practice, one should include those terms 
that give rise to the maximal stability of the results. 
In some sense, the variant ratio method of the previous 
section corresponds to considering the leading singular behavior and 
the first analytic correction---and also the leading nonanalytic term
if we further extrapolate the sequence.

On the cubic lattice, the large-order behavior is dictated by the 
singularities at $\pm \beta_c$. Indeed, given an observable $S$ 
with expansion $S = \sum_n c_n \beta^n$, for $n\to \infty$ the 
expansion coefficients behave like
\begin{eqnarray}
\beta_c^n c_n =&&  
n^{\zeta-1} 
\left( A_0 + {A_1 \over n^\Delta}  + {A_2 \over n} + 
{A_3 \over n^{1+\Delta}} + 
{A_4\over n^{2}} + ... \right)   \label{cnexp} \\
&&+(-1)^n n^{-(\theta_{\rm af}+1)} \left( B_0 +  {B_1\over n} + 
{B_2 \over n^{2}} + {B_{\Delta_{\rm af}} \over n^{\Delta_{\rm af}}} + ... 
\right). \nonumber
\end{eqnarray}
Here, we have neglected all subleading exponents except the first one
$\Delta$, and in particular, the first subdominant $\Delta_2$. 
However, since $\Delta_2 \approx 1$ \cite{NR-84}, for all practical purposes
a term $n^{-\Delta_2}$ cannot be distinguished from a purely analytic 
correction. Also, we do not write terms of order 
$n^{-k\Delta}$ since they cannot be 
distinguished from the analytic terms and corrections of order $n^{-m-\Delta}$.
Note also the presence of the parity-dependent corrections 
with exponent $\theta_{\rm af}$ and the subleading corrections 
with exponent $\Delta_{\rm af}$. For the susceptibility $\chi$, 
it is known \cite{Fisher-62} that $\theta_{\rm af} = 1-\alpha$. 
The argument can be generalized to all moments $m_{2k}$ and thus in 
all cases we predict $\theta_{\rm af} = 1-\alpha$.
We have tested this prediction for $\chi$, cf. Sec. \ref{iaan}, 
$m_2$, and $m_4$. By analyzing the expansion of $m_2$ with biased 
IA$k$'s that have a pair of singularities in $\pm \beta_c$,
 we obtain $\theta_{\rm af}=0.884(12)$, 
while from the expansion of $m_4$ we obtain $\theta_{\rm af}=0.90(9)$.
These results are clearly compatible with the prediction 
$\theta_{\rm af}=1-\alpha= 0.8904(5)$. For the exponent $\Delta_{\rm af}$
nothing is known. We have analyzed the expansion using $\Delta_{\rm af} = 1/2$
and $\Delta_{\rm af} = 1$. The results appear to be quite insensitive 
on either choice. For this reason, in the following we only report 
the results corresponding to purely analytic corrections, i.e., 
we set $B_{\Delta_{\rm af}}=0$.

Note that this method allows to determine the nonuniversal amplitudes
$A_0$, $A_\Delta$, $\ldots$, and consequently the amplitudes $a_i$ 
appearing in the expansion of $S$ for $\beta\to\beta_c$. If
\begin{equation}
S = A_S (\beta_c - \beta)^{-\zeta} \ \left[ 1 + a_S \left(\beta_c -
\beta\right)^\Delta\right] 
\end{equation}
then
\begin{eqnarray}
A_S &=& \Gamma(\zeta) A_0, \\
a_S &=& { \Gamma(\zeta-\Delta) A_1 \over \Gamma(\zeta) A_0}.
\label{cchi}
\end{eqnarray}
In the following, we shall perform two different analyses: 
(essentially) unbiased analyses in order to determine the 
exponents $\zeta$ and $\beta_c$ and biased analyses in which 
$\zeta$ and $\beta_c$ are fixed. In all cases we fix the value of 
$\Delta$, $\Delta = 0.52$, and the exponent of the antiferromagnetic 
singularity. In the unbiased analyses, in order to have a linear problem,
we consider $\ln c_n$ that behaves
\begin{eqnarray}
&&\ln c_n =  - {\rm ln} (\beta_c)\;n   + (\zeta-1) 
{\rm ln}\, n + b_0 + {b_1\over n^\Delta} +
{b_2\over n} + {b_3\over n^{1+\Delta}} +  {b_4\over n^2} 
+ ...  \label{lcnexp} \\
&&\qquad 
+(-1)^n n^{-(\zeta+\theta_{\rm af})} \left( d_0 +  {d_1\over n^\Delta} + 
{d_2 \over n} + {d_3\over n^{1+\Delta} } + 
{d_4\over n^2} ...  \right). \nonumber
\end{eqnarray}
As before, we have neglected terms that have exponents 
similar to those already present: for instance, terms $O(n^{-k\Delta - m})$
or $O(n^{-k\Delta_2 - h\Delta - m})$. In the expansion of the 
antiferromagnetic part we have assumed $\Delta_{\rm af} = \Delta$,
or $\Delta_{\rm af} = 1$. Note that if only analytic terms are present 
in Eq. (\ref{cnexp}), i.e., $B_{\rm af} = 0$, then $d_1$ is proportional
to $A_1$ and therefore it vanishes in improved models.

\begin{figure}[tbp]
\hspace{-1cm}
\vspace{0cm}
\centerline{\psfig{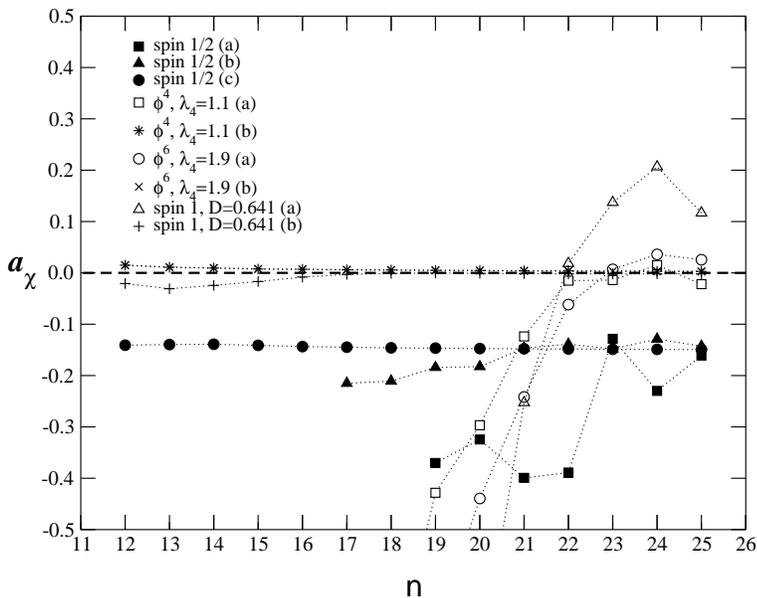}}
\vspace{0cm}
\caption{
Amplitude $a_\chi$ of the leading scaling correction
as obtained from several different analyses of $\chi$ for the standard spin-1/2
Ising model and for the three improved models.
Details are explained in the text.
}
\label{cdelta}
\end{figure}

We first analyze improved models and we verify that 
$A_1 \approx 0$. For this purpose, we consider the 
susceptibility $\chi$ and, for each improved Hamiltonian,
we generate two sequences of amplitudes in the following way:
\begin{itemize}
\item[(a)] We choose two integers $h,k$ and consider Eq. (\ref{lcnexp})
keeping only $b_0$, $\ldots$, $b_{h-1}$ in the ferromagnetic part
and $d_0$, $\ldots$, $d_{k-1}$ in the antiferromagnetic one. 
Then, we generate sequences $\beta_c^{(n)}$, $\gamma^{(n)}$,
$b_0^{(n)}$, $\ldots$, $b_{h-1}^{(n)}$, $d_0^{(n)}$, $\ldots$, $d_{k-1}^{(n)}$,
by solving the $(h+k+2)$ equations $\ln c_{n-m} = R_{n-m}$, 
$m=0,\ldots,h+k+1$, where $R_n$ is the right-hand side of Eq. (\ref{lcnexp}).
We use $\Delta = 0.52$, $\gamma + \theta_{\rm af} = 2.1277$.
\item[(b)] We 
choose two integers $h,k$ and consider Eq. (\ref{cnexp})
keeping only $A_0$, $\ldots$, $A_{h-1}$ in the ferromagnetic part
and $B_0$, $\ldots$, $B_{k-1}$ in the antiferromagnetic one.
We use $\Delta = 0.52$, $\gamma = 1.2373$, $\theta_{\rm af} = 0.8904$,
the IA estimate of $\beta_c$, and $B_{\rm af} = 0$.
Then, we generate sequences
$A_0^{(n)}$, $\ldots$, $A_{h-1}^{(n)}$, $B_0^{(n)}$, $\ldots$, $B_{k-1}^{(n)}$,
by solving the $(h+k)$ equations $c_{n-m} = R_{n-m}$,
$m=0,\ldots,h+k-1$, where $R_n$ is the right-hand side of Eq. (\ref{cnexp}).
\end{itemize}
In both cases we vary $h$ and $k$, trying to find the values that give 
the best stability of the exponents or of the leading amplitudes.
In the unbiased analysis (a), the preferred choice is 
$(h,k) = (4,4)$, while for analysis (b) we use $(h,k) = (3,2)$.
For these choices of the parameters,
in Fig. \ref{cdelta} we report the corresponding sequence of 
$a_\chi^{(n)} \equiv a_1^{(n)}$,
obtained using Eq. (\ref{cchi}). In the unbiased analysis (a), 
$a_\chi^{(n)}$ clearly converges to zero for the improved
Hamiltonians $\phi^4$ and $\phi^6$, as expected. For the spin-1 model, 
it is not that clear, and presumably more orders are need to observe
convincingly $a_\chi = 0$. In the case of the biased analysis,
$a_\chi^{(n)}$ is very stable and small already for 
$n\gtrsim 15$. For all Hamiltonians we observe 
$|a_\chi| \lesssim 10^{-3}$. 

As a second check of consistency we have verified that our estimates of 
$a_\chi$ are compatible with the quoted error bars on $\lambda_4^*$ and 
$D^*$. For this purpose, using the analysis of type (b) reported above,
we have computed $a_\chi$ for $\lambda_4^*\pm \Delta\lambda_4$, where 
$\Delta\lambda_4$ is the quoted error bar---for the spin-1 model 
we are referring to $D^*\pm \Delta D$. In all cases, we find 
$|a_\chi (\lambda_4^*\pm \Delta\lambda_4)| > |a_\chi (\lambda_4^*)|$ and 
that $a_\chi (\lambda_4^* + \Delta\lambda_4)$ and 
$a_\chi (\lambda_4^* - \Delta\lambda_4)$ have opposite sign. 
This confirms the correctness of our estimates of 
$\lambda_4^*$ and $D^*$. 
Of course, since we use $\beta_c$ and $\gamma$ obtained in the IA analysis,
the above results  represent only a check of consistency. Indeed:
(i) we determine $\beta_c$ and $\gamma$ by performing a IA analysis
whose results should be reliable only if the models are improved (in some 
sense we assume weakly here $a_\chi \approx 0$);
(ii) using such values of $\beta_c$ and $\gamma$, we estimate
$a_\chi$ and find $a_\chi\approx 0$.

\begin{figure}[tbp]
\hspace{-1cm}
\vspace{0cm}
\centerline{\psfig{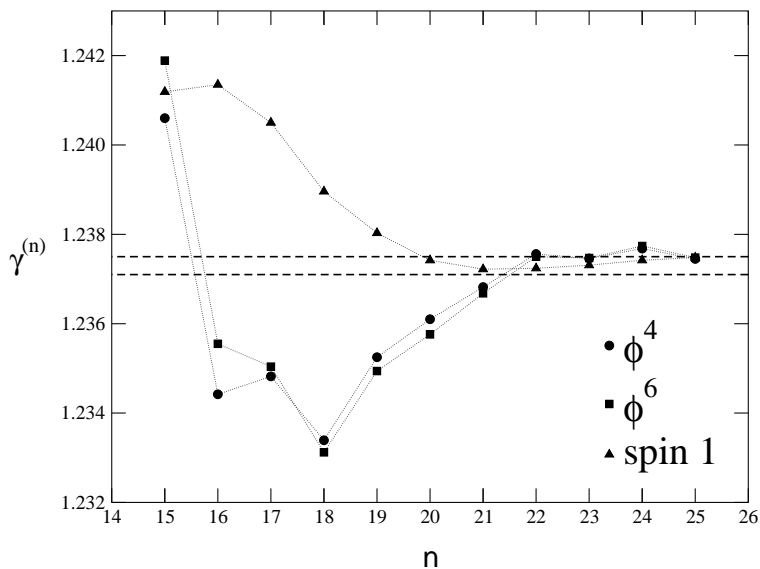}}
\vspace{0cm}
\caption{
Exponent $\gamma$ 
as obtained from the analysis (a) of $\chi$
for the three improved models.
Details are explained in the text.
}
\label{gammaGT}
\end{figure}

Once we have verified that $A_1$ is very small and compatible with zero 
within the precision of the analysis, we have performed 
several analyses fixing $A_1 = 0$ and $b_1=0$. At the same time, we have 
set $d_1 = 0$, which corresponds to assuming $\Delta_{\rm af} = 1$. 
We have determined the exponents by performing the analysis (a)
reported above.
In the case of the $\phi^4$ model for $\lambda_4=1.10$, this analysis gives
$\gamma\approx 1.2374$. 
Similarly, we obtain $\gamma\approx 1.2375$ for the $\phi^6$ model
at $\lambda_4=1.90$ and for the spin-1 model at $D=0.641$.
In Fig.~\ref{gammaGT} we show the sequence $\gamma^{(n)}$ 
for $(h,k)= (5,5)$ (since two coefficients vanish, we are considering 
four amplitudes in the ferromagnetic and antiferromagnetic expansion).
We observe a very good agreement with the IA estimate 
$\gamma = 1.2373(2)$.
It is difficult to estimate the uncertainty, 
since the results do not show a sufficiently robust 
stability with respect to the number $(h,k)$ of 
coefficients used in the analysis.

Finally, we report the
estimates of the amplitudes obtained in the analysis of type (b)
for the magnetic susceptibility:
\begin{itemize}
\item[$\phi^4$]: $A_0^{(\chi)}\approx 0.5246$, $A_2^{(\chi)} \approx 0.13$, 
$B_0^{(\chi)} \approx -0.0351$;
\item[$\phi^6$]: $A_0^{(\chi)}\approx 0.4601$, $A_2^{(\chi)} \approx 0.11$,
$B_0^{(\chi)}\approx -0.0311$;
\item[spin-1]: $A_0^{(\chi)}\approx 0.5126$, $A_2^{(\chi)} \approx 0.12$,
$B_0^{(\chi)} \approx -0.0359$.
\end{itemize}
Moreover, $|A_3^{(\chi)}|\lesssim 10^{-2}$ for the $\phi^4$ and 
$\phi^6$ models, while $A_3^{(\chi)}\approx -0.02$ for the spin-1
model. Errors should be $\pm 1$ on the last reported digit 
and include the uncertainty on the $n\to\infty$ extrapolation 
of the sequences and the variation of the estimates for 
$h$ and $k$ in the range $h=3-5$ and $k=2-4$. 
They do not take into account instead the variation of the estimates 
with $\gamma$ and $\beta_c$. Note that the estimate of $A_2$ 
is purely phenomenological and in practice it should correspond to the 
sum of the amplitude of $n^{-1}$ and of $n^{-\Delta_2}$ 
(note that in improved models the amplitude of $n^{-2\Delta}$ vanishes).

We have performed similar analyses for the spin-1/2 Ising model, in order to 
compute the nonuniversal amplitudes. We have 
performed: (a) an analysis of type (a) using $(h,k)=(4,4)$;
(b) an analysis of type (a) in which we have fixed $\beta_c$ 
to its MC value using $(h,k)=(4,3)$;
(c) an analysis of type (b) using $(h,k)=(3,2)$. The results 
for $a_\chi^{(n)}$ are reported in Fig.~\ref{cdelta}. 
These analyses give perfectly consistent results and allow us to 
determine the amplitudes: 
\begin{eqnarray}
A_0^{(\chi)} &=& 1.233 - 10 (\gamma-1.2373) - 0.013 (\Delta-0.52), \\
A_1^{(\chi)} &=& -0.13  - 0.7 (\Delta-0.52) + 50 (\gamma-1.2373), \\
B_0^{(\chi)} &=& -0.073,
\end{eqnarray}
where we have explicitly written the dependence on the input
parameters (when it turns out to be relevant).
We have repeated the same analysis for the second moment $m_2$.
We obtain
\begin{eqnarray}  
A_0^{(m_2)}&=& 1.301 - 10 (\gamma+ 2\nu - 2.49754) - 0.07 (\Delta-0.52), \\
A_1^{(m_2)} &=& -0.73 - 4(\Delta-0.52) + 55 (\gamma+2\nu-2.49754),\\
B_0^{(m_2)} &=& 0.06.
\end{eqnarray}
Using the above  results and Eq.~(\ref{cchi}), one can determine 
the amplitudes $a_\chi$ and  $a_\xi$, associated with the $O(t^\Delta)$
of the scaling corrections in the Wegner expansion of $\chi$ and $\xi$
respectively, and evaluate their universal ratio.
We obtain $a_\xi/a_\chi=0.9(1)$, where the error takes also into account
the uncertainty on the input parameters of the biased analysis.
For comparison we mention the recent HT result 
$a_\xi/a_\chi=0.76(6)$ \cite{BC-02}, and the
field theoretical estimate 
$a_\xi/a_\chi=0.68(2)$ \cite{BB-85}.

\section{The critical equation of state}
\label{CES}

\subsection{Definitions} \label{CES.1}

The equation of state relates the 
magnetization $M$, the magnetic field $H$, and the reduced temperature 
$t\equiv (T-T_c)/T_c$. 
In the neighborhood of the critical point $t=0$, $H=0$, it can be 
written in the scaling form 
\begin{eqnarray}
&&H = B_c^{-\delta}M^{\delta} f(x), \\
&&x \equiv t (M/B)^{-1/\beta},
\label{eqstfx}
\end{eqnarray}
where $B_c$ and $B$ are the amplitudes of the magnetization on the critical 
isotherm and on the coexistence curve,
\begin{eqnarray}
    M &=& B_c H^{1/\delta}\qquad\qquad t=0, \label{def-Bcconst} \\
    M &=& B (-t)^\beta\qquad\qquad  H=0,\,\, t<0. \label{def-Bconst}
\end{eqnarray}
Using these normalizations the coexistence curve corresponds to $x=-1$,
and the universal function $f(x)$ satisfies $f(-1)=0$, $f(0)=1$.
Griffiths' analyticity implies that $f(x)$ is regular everywhere for $x>-1$. 
It has a regular expansion in powers of $x$,
\begin{equation}
f(x) = 1 + \sum_{n=1}^\infty f_n^0 x^n,
\label{expansionfx-xeq0}
\end{equation}
and a large-$x$ expansion of the form
\begin{equation}
f(x) = x^\gamma \sum_{n=0}^\infty f_n^\infty x^{-2n\beta}.
\label{largexfx}
\end{equation}
At the coexistence curve, i.e., for $x\rightarrow -1$,
$f(x)$ has at most an essential singularity \cite{singatcoexcurve}.
It can be asymptotically expanded as
\begin{equation}
f(x) \approx  \sum_{n=1}^\infty f^{\rm coex}_n (1+x)^n. 
\label{fxcc} 
\end{equation}

It is useful to rewrite the equation of state in terms of a variable 
proportional to $M t^{-\beta}$, although in this case we must distinguish 
between $t > 0$ and $t < 0$. For $t > 0$ we define
\begin{eqnarray}
&& H = \left({C^+\over C_4^+}\right)^{1/2} t^{\beta \delta} F(z),
\nonumber \\
&& z \equiv \left[- {C_4^+\over (C^+)^3} \right]^{1/2} M t^{-\beta},
\label{defFz}
\end{eqnarray}
while for $t<0$ we set
\begin{eqnarray}
&& H = {B\over C^-} (-t)^{\beta \delta} \Phi(u), 
\nonumber \\ 
&& u \equiv {M\over B} (-t)^{-\beta}.
\label{defPhiu}
\end{eqnarray}
The constants $C^\pm$ and $C^+_4$ are the amplitudes appearing in the 
critical behavior 
of the zero-momentum connected $n$-point correlation functions $\chi_n$:
\begin{equation}
\chi_n = C_n^\pm |t|^{-\gamma - (n-2) \beta\delta}.
\label{defCn}
\end{equation}
The susceptibility $\chi$ corresponds to $\chi_2$ and we simply write 
$C^\pm = C_2^\pm$.

With the chosen normalizations \cite{GZ-97,PV-98,PV-99}
\begin{eqnarray}
F(z) &=&
z + {1\over 6} z^3
       + \sum_{j=3} {1\over (2j-1)!} r_{2j}\, z^{2j-1},
\label{Fzexpa} \\
\Phi(u) &=& (u-1) + \sum_{j=3}^\infty {1\over (j-1)!} v_{j}\, (u-1)^{j-1}.
\label{phiuexp}
\end{eqnarray}
The functions $F(z)$ and $\Phi(u)$ are related to $f(x)$.
Indeed,
\begin{equation}
 z^{-\delta} F(z) = F_0^\infty f(x), \qquad\qquad  z = z_0
 x^{-\beta}, 
\end{equation}
and
\begin{equation}
 u^{-\delta} \Phi(u) = {C^- B^{\delta-1}\over B_c^\delta} f(x),
     \qquad\qquad  u = (-x)^{-\beta}.
\label{rel-Phiu-fx}
\end{equation}
The constant $F_0^\infty$ is defined by the large-$z$ behavior of $F(z)$,
\begin{equation}
F(z) = z^\delta \sum_{k=0} F^{\infty}_k z^{-k/\beta},
\label{asyFz}
\end{equation}
while
\begin{equation}
z_0 = \left[- {C_4^+\over (C^+)^3} \right]^{1/2} B.
\end{equation}
To compare with experimental data, it is useful to determine the 
magnetization as a function of $t H^{-1/\beta\delta}$.
Therefore, we define 
\begin{eqnarray}
&&E(y) \equiv B_c^{-1} M H^{-1/\delta}  =  f(x)^{-1/\delta}  ,\\
&& y \equiv     (B/B_c)^{1/\beta} t H^{-1/(\beta\delta)} = 
x f(x)^{-1/(\beta\delta)}.
\label{Ey} 
\end{eqnarray}
Finally, we shall also determine the scaling behavior of the susceptibility,
by defining 
\begin{equation}
D(y) \equiv   B_c^{-1} H^{1-1/\delta} \chi =
 { f(x)^{1-1/\delta} \over \delta f(x) - {1\over\beta} x f'(x)}.
\end{equation}

\subsection{Small-magnetization behavior} \label{IHTrj}

In this section we determine the first few 
coefficients $r_{2j}$ appearing in the expansion of 
the scaling function $F(z)$, cf. Eq. (\ref{defFz}).
We shall also compute the four-point renormalized coupling constant
$g_4$, which, although not related to the equation of state,
is relevant for the field-theoretical approach and will be used
to determine amplitude ratios involving the second-moment correlation
length.

In order to estimate 
the critical limit of $g_4$ and of 
$r_{2j}$ we first determine their HT expansions using the 
corresponding results for $\chi_{2j}$ and $m_2$ 
\begin{eqnarray}
&&g_4 = - {\chi_4\over \chi^2 \xi^3}, \label{g4def}\\
&&r_6 = 10 - {\chi_6\chi_2\over \chi_4^2},\label{r6gr}\\
&&r_8 = 280 - 56{\chi_6\chi_2\over \chi_4^2} + 
        {\chi_8\chi_2^2\over \chi_4^3}.\label{r8gr}
\end{eqnarray}
The corresponding series \cite{foot-series} have been analyzed 
following closely the procedure presented in App. B.3 of 
Ref.~\cite{CHPRV-01}.
We use biased IA1's with a singularity at $\beta_c$ or 
a pair of singularities at $\pm\beta_c$, where $\beta_c$ is 
obtained from the analysis of the susceptibility.
Around $\beta_c$, IA1's behave like \cite{int-appr-ref}
\begin{equation}
{\rm IA1} \approx 
f(\beta) \left(1 - \beta/\beta_c\right)^{\zeta} + g(\beta),
\label{IA1bh}
\end{equation}
where $f(\beta)$ and $g(\beta)$ are regular at $\beta_c$, provided
$\zeta$ is not a negative integer.  In particular 
\begin{equation}
\zeta = {P_0(\beta_c)\over P_1'(\beta_c)},\qquad\qquad 
    g(\beta_c) = - {R(\beta_c)\over P_0(\beta_c)}
\label{IA1bhf}
\end{equation}
(see Eq.~(\ref{IAkdef}) for the definition of the above quantities).
In the case we are considering, $\zeta$ is positive and, therefore,
$g(\beta_c)$ provides the desired  estimate.  

In Table~\ref{grr2j} (first line) we report the estimates of  $g_4$ obtained
for the three improved Hamiltonians. 
The error in parentheses is related to the spread of the 
approximants and the
second one in brackets to the uncertainty on $\lambda_4^*,D^*$.
The error induced by the uncertainty on $\beta_c$ is negligible.
The results  are perfectly consistent.
Our final estimate is
\begin{equation}
g_4=23.56(2).
\end{equation}
The result for the exponent
$\zeta$ in Eq.~(\ref{IA1bh}) is $\zeta=1.3(3)$, which is consistent
with  our expectation 
for improved models, i.e.,
$\zeta =\Delta_2 \approx 2 \Delta$ and $\Delta \approx 0.5$.
For comparison, the same analysis applied to the standard
Ising model gives $g_4=23.5(5)$ and $\zeta=0.6(3)$,
in agreement with the fact that in this case $\zeta=\Delta$.
Notice that the small difference with the estimate of $g_4$ reported
in Ref.~\cite{CPRV-99} is essentially due to the different 
analysis employed here, which is better justified due to the
nonanalytic behavior at $\beta_c$  predicted by renormalization group
 \cite{CPRVan}.
With respect to standard Pad\'e approximants,
biased IA1's  require more terms of the series to 
give reasonable results, but they are  less subject to
systematic errors since they allow for confluent 
nonanalytic corrections at $\beta_c$.
Biased IA1's give \cite{PV-review}
$g_4=23.54(4)$ when applied to the 17th-order series
of Ref.\cite{CPRV-99}.

Results for  $r_{6}, r_8$ are obtained using the same method
and are reported in Table~\ref{grr2j}. 
We finally recall that a rough estimate of
$r_{10}$ was obtained  in Ref.~\cite{CPRV-99} from the analysis
of its 15th-order series, obtaining $r_{10}=-13(4)$.
A review of the available results for these quantities can be found
in Ref.~\cite{PV-review}.

\begin{table}[tbp]
\footnotesize
\begin{center}
\caption{
Results for $g_4$, $r_6$ and $r_8$.
}
\label{grr2j}
\begin{tabular}{ccccc}
\hline\hline
\multicolumn{1}{l}{}&
\multicolumn{1}{c}{$\phi^4$}&
\multicolumn{1}{c}{$\phi^6$}&
\multicolumn{1}{c}{spin-1}&
\multicolumn{1}{c}{final estimates}\\
\hline \hline
$g_4$ & 23.559(8)[11] & 23.554(8)[20] & 23.560(20)[5] &  23.56(2) \\ 

$r_6$ & 2.057(4)[1] & 2.056(4)[2] & 2.052(8)[2] & 2.056(5) \\

$r_8$ & 2.29(9)[3] & 2.31(5)[5] & 2.37(7)[3] & 2.3(1) \\
\hline\hline
\end{tabular}
\end{center}
\end{table}

\subsection{Parametric representations of the equation of state}
\label{appeq}

In this section we shall determine the equation of state 
using parametric representations, improving  the 
results of Refs. \cite{GZ-97,CPRV-99}. This method has also been 
applied in two dimensions \cite{CHPV-01}, and
to the three-dimensional $XY$ \cite{CPRV-00-2,CHPRV-01}
and Heisenberg \cite{CHPRV-02} universality classes.

In order to obtain approximate expressions for the equation of state,
we parametrize the thermodynamic variables in terms of two parameters
$R$ and $\theta$, implementing all expected
scaling and analytic properties. Explicitly, we write 
\cite{Schofield-69,SLH-69,Josephson-69} 
\begin{eqnarray}
M &=& m_0 R^\beta \theta ,\nonumber \\
t &=& R(1-\theta^2), \nonumber \\
H &=& h_0 R^{\beta\delta}h(\theta), \label{parrep}
\end{eqnarray}
where $h_0$ and $m_0$ are normalization constants.  
The function
$h(\theta)$ is odd and normalized so that 
$h(\theta)=\theta+O(\theta^3)$.
The smallest positive zero of $h(\theta)$, which should satisfy
$\theta_0>1$, corresponds to the coexistence curve, i.e., to $T<T_c$
and $H\to 0$. 
We mention that alternative versions of the parametric representations
have been considered in Ref.~\cite{otherparam}.

It is easy to express the scaling functions introduced in Sec. \ref{CES.1}
in terms of $\theta$.
The scaling function $f(x)$ is obtained from
\begin{eqnarray}
&& x = {1 - \theta^2\over \theta_0^2 - 1} 
\left( {\theta_0\over \theta}\right)^{1/\beta}, \nonumber \\
&& f(x) = \theta^{-\delta} {h(\theta)\over h(1)},
\label{fxmt}
\end{eqnarray}
while $F(z)$ is obtained by
\begin{eqnarray}
&&z = \rho \theta \left( 1 - \theta^2\right)^{-\beta},
\nonumber \\
&&F(z(\theta)) = \rho \left( 1 - \theta^2 \right)^{-\beta\delta} h(\theta),
\label{Fzrel}
\end{eqnarray}
where $\rho$ can be related to $m_0$, $h_0$, $C^+$ and $C_4^+$ using 
Eqs. (\ref{defFz}) and (\ref{parrep}).

It is important to note that Eq. (\ref{parrep}) and the normalization
condition $h(\theta)\approx \theta$ for $\theta\to 0$ do not completely
fix the function $h(\theta)$. Indeed, one can rewrite
the relation between $x$ and $\theta$ in the form
\begin{equation}
x^\gamma = h(1) \, f_0^\infty (1 - \theta^2)^\gamma 
    \theta^{1-\delta}.
\end{equation}
Thus, given $f(x)$,
the value of $h(1)$ can be arbitrarily chosen to 
completely fix $h(\theta)$. In the expression (\ref{Fzrel}) 
we can fix this arbitrariness by choosing arbitrarily the parameter $\rho$
\cite{GZ-97,CPRV-99,CPRV-00-2,PV-review}.  

As suggested by 
arguments based on the $\epsilon$-expansion \cite{GZ-97,CPRV-99},
we approximate $h(\theta)$ with polynomials, i.e., we set
\begin{equation} 
     h(\theta) = \theta + \sum_{n=1}^k h_{2n+1} \theta^{2n+1}.
\label{parametrich}
\end{equation} 
This choice is further supported by the effectiveness of its
simplest version with $k=1$, which is the so-called linear model.
If we require   
the approximate parametric representation 
to give the correct $(k-1)$ universal ratios $r_6$, $r_8$, 
$\ldots$, $r_{2k+2}$, we obtain
\begin{equation}
   h_{2n+1} = \sum_{m=0}^n c_{nm} 6^m (h_3 + \gamma)^m {r_{2m+2}\over (2m+1)!},
\label{hcoeff}
\end{equation}
where
\begin{equation}
   c_{nm} = {1\over (n-m)!} \prod_{k=1}^{n-m} 
   (2\beta m - \gamma + k - 1),
\end{equation}
and we have set $r_2 = r_4 = 1$. 
Moreover, by requiring that $F(z)=z + {1\over 6} z^3 + ...$,
we obtain the relation
\begin{equation}
   \rho^2 = 6 (h_3 + \gamma).
\label{h3intermsofrho}
\end{equation}
In the exact parametric representation, the coefficient $h_3$ 
can be chosen arbitrarily. Of course, 
this is no longer true when we use our truncated function $h(\theta)$, 
and the related approximate function  $f_{\rm approx}^{(k)}(x,h_3)$ 
depends on $h_3$. We must thus fix a particular value for this parameter. 
Here we use a variational approach, requiring the approximate function
$f_{\rm approx}^{(k)}(x,h_3)$ to have the smallest possible dependence 
on $h_3$.
Thus, we set $h_3 = h_{3,k}$, where $h_{3,k}$ is 
a solution of the global stationarity condition
\begin{equation}
  \left. 
  {\partial f_{\rm approx}^{(k)}(x,h_3) \over 
   \partial h_3}\right|_{h_3 = h_{3,k}} 
  = 0
\label{globalstationarity}
\end{equation}
for all $x$. Equivalently one may require that, 
for {\em any} universal ratio $R$ that 
can be obtained from the equation of state, its approximate 
expression $R_{\rm approx}^{(k)}$ obtained by using the 
parametric representation satisfies 
\begin{equation}
  \left.
  {d R_{\rm approx}^{(k)}(h_3) \over d h_3}\right|_{h_3 = h_{3,k}}
  = 0.
\label{globalstationarity2}
\end{equation}
The existence of such a value of $h_3$ is a nontrivial mathematical fact.
The stationary value of $h_3$ is the solution of the 
algebraic equation \cite{CPRV-99}
\begin{equation}
\left[ 2 (2\beta - 1) (h_3 + \gamma) {\partial\over \partial h_3} - 2 \gamma 
   + 2 k\right] h_{2k + 1} = 0.
\label{stationaritycondition}
\end{equation}
For $k=1$, the so-called linear model, Eq. (\ref{stationaritycondition})
gives
\begin{equation}
    h_3 = {\gamma (1 - 2 \beta)\over \gamma - 2 \beta},
\end{equation}
which is the optimal value of $h_3$ considered in Ref.~\cite{SLH-69}.
Thus, the optimal (sometimes called restricted)
linear model represents the first
approximation of our scheme.

\subsection{Results}
\label{reseq}

\begin{table*}
\footnotesize
\begin{center}
\caption{
Polynomial approximations of $h(\theta)$ using the global stationarity
condition for various values of the parameter $k$.
The reported expressions are obtained by using the central values
of the input parameters. The last column shows
the corrections to the simple linear model $h_{\rm lin}(\theta,\theta_0)\equiv 
\theta ( 1 - \theta^2/\theta_0^2)$.
}
\label{trht}
\begin{tabular}{clcl}
\hline\hline
$k$ &  $\qquad\qquad\quad  h(\theta)/\theta$ 
& $\theta_0^2$ 
& $\qquad h(\theta)/h_{\rm lin}(\theta,\theta_0)$  \\
\hline\hline
1  &  $ 1 - 0.734732 \theta^2 $ & $\quad$ 1.36104 $\quad$   & 1 \\
2  &  $ 1 - 0.731630 \theta^2 + 0.009090 \theta^4 $ &  $\quad$ 1.39085
$\quad$ &  
$1 - 0.0126429 \theta^2$ \\
3  &  
  $ 1 - 0.736743 \theta^2 + 0.008904 \theta^4 - 0.000472 \theta^6 $ & 
$\quad$  1.37861 $\quad$  &    $1-0.0113775 \theta^2 + 0.0006511 \theta^4$ \\
\hline\hline
\end{tabular}
\end{center}
\end{table*}

\begin{figure}[tbp]
\hspace{-1cm}
\vspace{0.2cm}
\centerline{\psfig{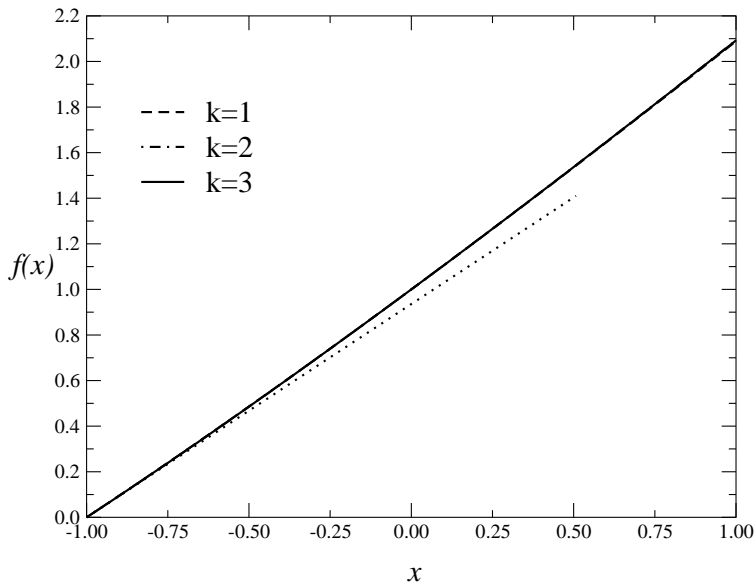}}
\vspace{-0.2cm}
\caption{
The scaling function $f(x)$. 
We also plot the asymptotic behavior of $f(x)$ at the coexistence
curve (dotted line), i.e., $f(x)\approx f_0^{\rm coex} (1+x)$
for $x\rightarrow -1$.}
\label{figfx}
\end{figure}

\begin{figure}[tbp]
\hspace{-1cm}
\vspace{0cm}
\centerline{\psfig{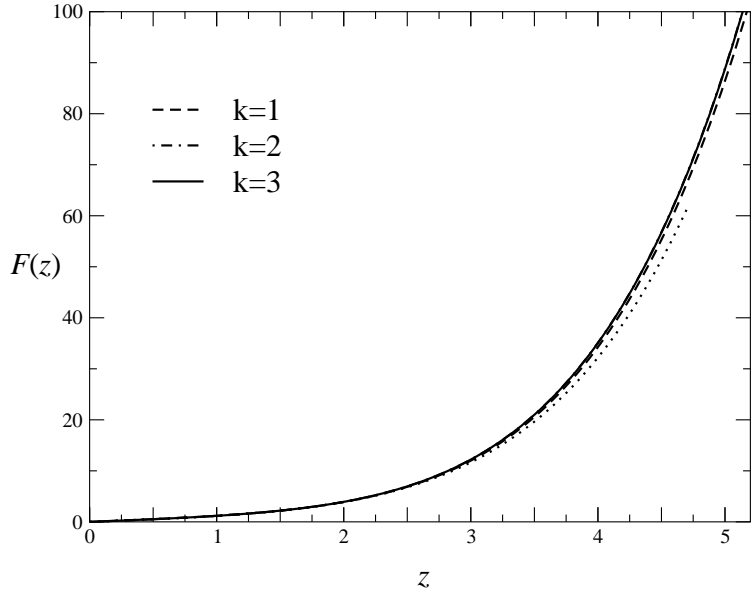}}
\vspace{0cm}
\caption{
The scaling function $F(z)$. 
We also show the plot of the small-$z$
expansion (dotted line), i.e., 
$F(z)\approx z + \frac{1}{6} z^3 + 
\frac{1}{120} r_6 z^5$ for $z\rightarrow 0$.
}
\label{figFz}
\end{figure}

\begin{figure}[tbp]
\hspace{-1cm}
\vspace{0cm}
\centerline{\psfig{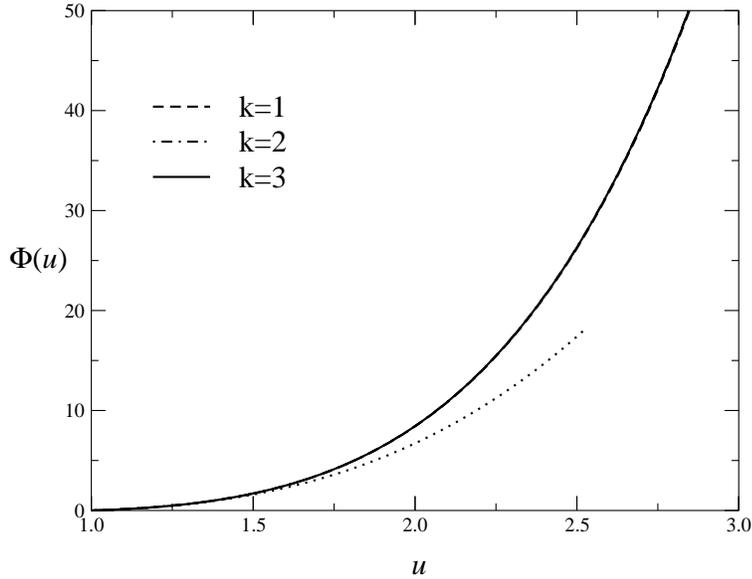}}
\vspace{0cm}
\caption{
The scaling function $\Phi(u)$. 
We plot also the asymptotic behavior of $\Phi(u)$ at the coexistence
curve (dotted line), i.e., $\Phi(u) \approx 
(u-1) + \frac{1}{2} v_3 (u-1)^2 + \frac{1}{6} v_4 (u-1)^3$
for $u\rightarrow 1$.
}
\label{figPhiu}
\end{figure}

Following  Ref.~\cite{CPRV-99}, we apply
the variational method by using the HT results for 
$\gamma=1.2373(2)$, 
$\nu=0.63012(16)$, 
$r_6=2.056(5)$, 
$r_8=2.3(1)$, and 
$r_{10}=-13(4)$ as input
parameters of the approximation scheme. This provides different
approximations with $k=1,2,3,4$. 
In Table~\ref{trht} we report
the polynomials $h(\theta)$ for $k=1,2,3$,
that are obtained in the variational approach 
for the central values of the input parameters.
The fast decrease of the coefficients of the higher-order terms in $h(\theta)$
gives further support to the effectiveness of the approximation scheme. 
We do not report  $h(\theta)$ for $k=4$, 
since it requires $r_{10}$ and its available estimate is rather imprecise.
Using the results reported in Table~\ref{trht} and Eqs.~(\ref{fxmt}),  
(\ref{Fzrel}), and (\ref{rel-Phiu-fx}),
one may easily compute the corresponding approximations
for the scaling functions $f(x)$, $F(z)$, and $\Phi(u)$.
The results show a good convergence with increasing $k$.  
Actually, the results 
for $k=2,3,4$ are already consistent within the errors induced by
the uncertainty on the input parameters, indicating that 
the systematic error due to the truncation is at most of the same order 
of the error induced by the input data. 
In  Figs.~\ref{figfx}, \ref{figFz}, and \ref{figPhiu} we show respectively 
the scaling functions as obtained from $h(\theta)$ for $k=1,2,3$.

In Table~\ref{eqstdet0} we report results concerning the behavior 
of the scaling function $f(x)$, $F(z)$ and $\Phi(u)$ for $H=0$ and on the 
critical isotherm, 
cf. Eqs. (\protect\ref{expansionfx-xeq0}), 
(\protect\ref{largexfx}), (\protect\ref{fxcc}), (\protect\ref{Fzexpa}), 
(\protect\ref{phiuexp}), (\protect\ref{asyFz}).
Note that
the results for $k=1,2,3$ oscillate and that the uncertainty due to the 
input parameters on the $k=3$ results is approximately the 
same as the difference between the estimates with $k=2$ and $k=3$.
Therefore, it is reasonable to consider the $k=3$ truncation as the 
best approximation of the method
using the available input parameters and to use the corresponding errors
as final uncertainties. 

\begin{table}[tbp]
\footnotesize
\begin{center}
\caption{
Expansion coefficients for the 
scaling equation of state obtained by the variational approach. 
See text for definitions.
Numbers marked
with an asterisk are inputs, not predictions.  
}
\label{eqstdet0}
\begin{tabular}{ccccc}
\hline\hline
\multicolumn{1}{c}{}&
\multicolumn{1}{c}{$k=1$}&
\multicolumn{1}{c}{$k=2$}&
\multicolumn{1}{c}{$k=3$}&
\multicolumn{1}{c}{$k=4$}\\
\hline \hline
$\theta_0^2$ & 1.3610(8) & 1.390(2) & 1.38(2) & 1.34(6) \\
$\rho$  & 1.7365(8) & 1.741(1) & 1.733(10) &  1.69(6) \\\hline

$r_6$ & 1.938(3) & $^*$2.056(5) &  $^*$2.056(5) &  $^*$2.056(5) \\

$r_8$ & 2.50(2) & 2.39(3) & $^*$2.3(1) & $^*$2.3(1) \\

$r_{10}$ & $-$12.59(2) & $-$12.08(5) &  $-$10.6(1.8) & $^*-$13(4)\\

$F^\infty_0$ & 0.03277(8) & 0.03388(11) & 0.03382(15)  & 0.0338(2) \\

$z_0$ & 2.8254(7) & 2.792(2) & 2.794(3) & 2.798(8) \\ 

$f^0_1$ & 1.05041(7) & 1.0532(2) & 1.0527(7) & 1.051(2) \\

$f^0_2$ & 0.04298(6) & 0.04494(13) &  0.0446(4) & 0.0439(13) \\

$f^0_3$ & $-$0.02474(4) & $-$0.02595(8) & $-$0.0254(7) & $-$0.023(4) \\

$f^\infty_0$ & 0.5960(4) & 0.6031(7)  & 0.6024(15)  & 0.601(4)  \\

$f_1^{\rm coex}$ & 0.93912(9) &  0.9347(3) & 0.9357(11) & 0.938(4) \\

$v_3$ & 6.013(4) & 6.062(4) & 6.050(13) & 6.02(5) \\

$v_4$ & 16.32(3) & 16.10(4) & 16.17(10)  & 16.4(3) \\
\hline\hline
\end{tabular}
\end{center}
\end{table}

\begin{figure}[tbp]
\hspace{-1cm}
\vspace{0.2cm}
\centerline{\psfig{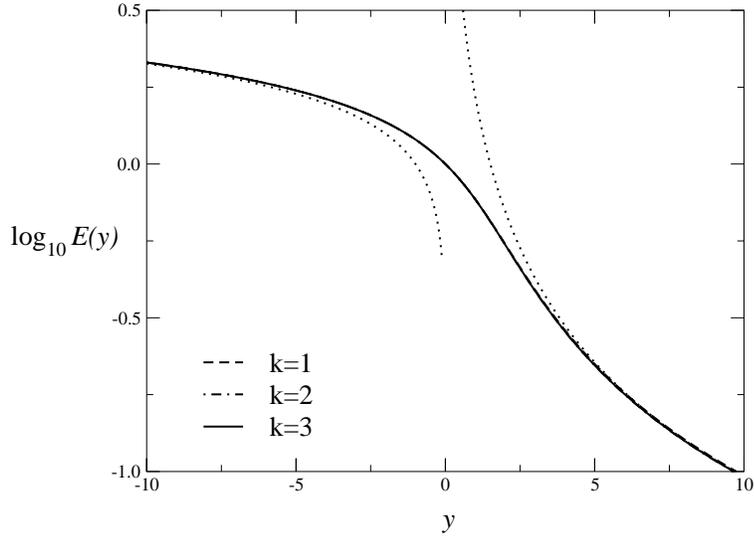}}
\vspace{-0.2cm}
\caption{
The scaling function $E(y)$. We also report its asymptotic behaviors
(dotted lines):
$E(y)\approx R_\chi y^{-\gamma}$ for $y\rightarrow +\infty$, and
$E(y)\approx - y^{\beta}$ for $y\rightarrow -\infty$.
}
\label{figEw}
\end{figure}

\begin{figure}[tbp]
\hspace{-1cm}
\vspace{0.2cm}
\centerline{\psfig{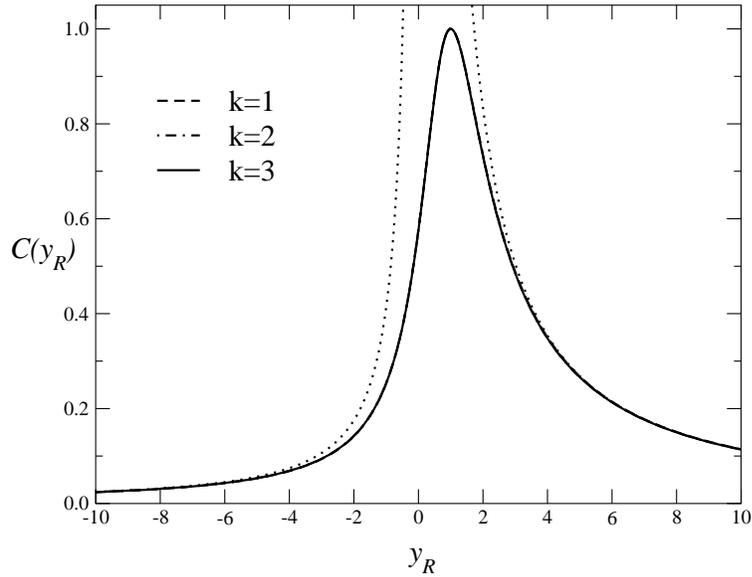}}
\vspace{-0.2cm}
\caption{
The scaling function $C(y_R)$. 
We also report its asymptotic behaviors (dotted lines):
$C(y_R)\approx R_\chi y_{\rm max}^{-\gamma} D(y_{\rm max})^{-1}
y_R^{-\gamma}\approx 1.97 y_R^{-\gamma}$ for $y_R\rightarrow +\infty$, and
$C(y_R)\approx 
\beta (f_1^{\rm coex})^{-1} y_{\rm max}^{-\gamma} D(y_{\rm max})^{-1}
(-y_R)^{-\gamma}\approx 0.413 (-y_R)^{-\gamma}$ for $y_R\rightarrow -\infty$.
}
\label{figCu}
\end{figure}

In Fig. \ref{figEw} we give the behavior of the
magnetization as a function of $t$ and
$H$, reporting the scaling function $E(y)$.
The behavior of the susceptibility can be obtained from the 
scaling function $D(y)$.
The function 
$D(y)$ has a maximum for $y_{\rm max} = 1.980(4)$, corresponding to the 
so-called crossover or pseudocritical line
(see Sec.~\ref{univratio}).
In order to simplify possible comparisons, it may be convenient to
consider the rescaled function
\begin{eqnarray}
&&C(y_R) = {D(y)\over D(y_{\rm max}) }, \nonumber \\
&& y_R = {y\over y_{\rm max}}, \label{Cwb}
\end{eqnarray}
which is such that 
the maximum corresponds to $y_R=1$ and satisfies $C(1) =1$.
In Fig.~\ref{figCu}  we plot the scaling function $C(y_R)$,
as obtained from the $k=1,2,3$ approximate parametric representations.

\begin{figure}[tb]
\hspace{-1cm}
\vspace{0cm}
\centerline{\psfig{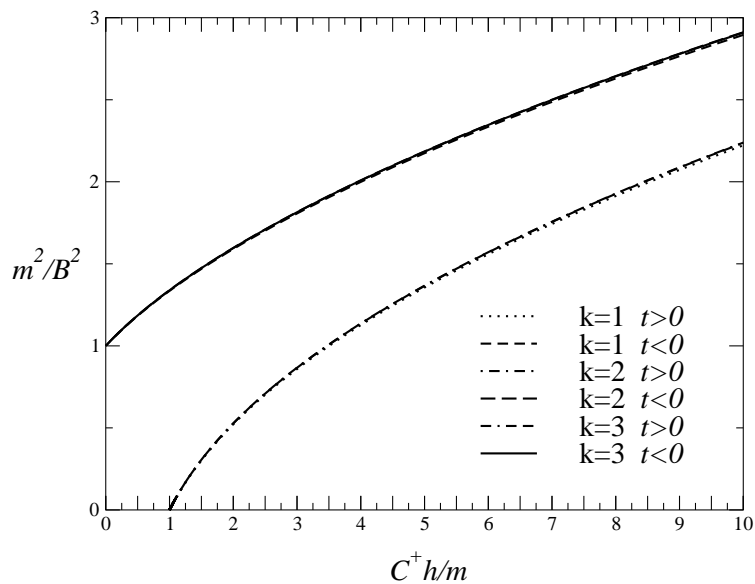}}
\vspace{0cm}
\caption{
Plot of $m^2/B^2$ versus $C^+ h/m$.
}
\label{m2vshsum}
\end{figure}

In experimental work on magnetic systems, it is customary to report
\cite{KR-67,BK-97}
$h/m\equiv H |t|^{-\gamma}/M$ versus $m^2 = M^2 |t|^{-2\beta}$. 
Such a function 
can be easily obtained from our approximations for $f(x)$, since 
$m^2 = B^2 |x|^{-2\beta}$ and 
\begin{equation} 
   {h\over m} = k |x|^{-\gamma} f(x),
\end{equation}
where the constant $k$ can be written as 
\begin{equation}
   k = B_c^{-\delta} B^{\gamma/\beta} = {R_\chi\over C^+},
\end{equation}
where $R_\chi \equiv C^+ B^{\delta-1}/ B_c^\delta$ is a universal constant,
see Sec. \ref{univratio}.
A plot of $m^2/B^2$ versus $C^+ h/m$ 
for the two cases $t>0$ and $t<0$ is reported in Fig. \ref{m2vshsum}.

It is interesting to observe that in a neighborhood of the critical
isotherm the equation of state can be written in the 
Arrott-Noakes form \cite{AN-67}
\begin{equation}
\left({H\over M}\right)^{1/\gamma} = a t + b M^{1/\beta},
\label{Arrott-Noakes}
\end{equation}
where $a$ and $b$ are numerical constants. Indeed, using the results 
of Table \ref{eqstdet0} for $k=3$, we obtain 
\begin{equation}
\left({H\over M}\right)^{1/\gamma} k^{-1/\gamma} =
\left({M\over B}\right)^{1/\beta} +
 0.851\ t - 0.050\, t^2 \left({M\over B}\right)^{-1/\beta}
            - 0.008\, t^3 \left({M\over B}\right)^{-2\beta}
        \cdots
\end{equation} 
Thus, corrections to Eq. (\ref{Arrott-Noakes}) are small and thus 
this expression has a quite wide range of validity.

\section{Universal amplitude ratios}
\label{univratio}

From the critical equation of state one may derive estimates
of several universal amplitude ratios.
They are expressed in terms of the amplitudes of the 
magnetization, cf. Eqs. (\ref{def-Bcconst}) and (\ref{def-Bconst}), 
of the magnetic susceptibility and $n$-point correlation functions,
cf. Eq. (\ref{defCn}), of the specific heat
\begin{equation}
C_{H} = A^{\pm} |t|^{-\alpha},
\end{equation}
of the second-moment correlation length 
\begin{equation}
\xi = f^{\pm} |t|^{-\nu},
\label{xiamp}
\end{equation}
and of the true (on-shell) correlation length, 
describing the large distance behavior of the two-point function,
\begin{equation}
\xi_{\rm gap} = f_{\rm gap}^{\pm} |t|^{-\nu}.
\label{xiosamp}
\end{equation}
One can also define amplitudes along the critical isotherm, e.g.\ 
\begin{eqnarray}
\chi &=& C^c |H|^{-{\gamma/\beta\delta}}, \label{chicris}\\
\xi &=& f^c |H|^{-{\nu/\beta\delta}}, \label{xicris}\\
\xi_{\rm gap} &=& f_{\rm gap}^c |H|^{-{\nu/\beta\delta}}. \label{xigapcris}
\end{eqnarray}
We also consider the crossover (or pseudocritical) line $t_{\rm max}(H)$, 
that is defined as the reduced temperature for which
the magnetic susceptibility has a maximum at $H$ fixed. 
Renormalization group predicts 
\begin{eqnarray}
&&t_{\rm max}(H) = T_p H^{1/(\gamma+\beta)},\\
&&\chi(t_{\rm max},H)= C_p t_{\rm max}^{-\gamma}.
\end{eqnarray}
We consider several universal amplitude ratios.
They are defined in Table~\ref{notationsur}.

\begin{table*}
\begin{center}
\begin{tabular}{|l|l|}
\hline
& \\[-1mm]
$U_0\equiv A^+/A^-$ & $U_2\equiv C^+/C^-$\\[1.6mm] 
$U_4\equiv C^+_4/C^-_4$ & $R_4^+\equiv - C_4^+B^2/(C^+)^3$ \\[1.6mm] 
$R_c^+\equiv \alpha A^+C^+/B^2$ &$R_c^-\equiv \alpha A^- C^-/B^2$ \\[1.6mm] 
$R_4^-\equiv C_4^-B^2/(C^-)^3$ & 
    $R_\chi\equiv C^+ B^{\delta-1}/(B_c)^\delta$ \\[1.6mm]
$v_3\equiv - C_3^-B/(C^-)^2$  & $v_4 \equiv -C_4^-B^2/(C^-)^3 + 3 v_3^2 $ \\
   [1.6mm]
$g_4^+\equiv -C_4^+/[ (C^+)^2 (f^+)^3] $ & 
    $w^2\equiv C^- /[ B^2 (f^-)^d]$ \\  [1.6mm]
$U_\xi\equiv f^+/f^-$ & 
   $U_{\xi_{\rm gap}}\equiv f^+_{\rm gap}/f^-_{\rm gap}$ \\  [1.6mm]
$Q^+ \equiv \alpha A^+ (f^+)^3$&$ Q^- \equiv \alpha A^- (f^-)^d $ \\ [1.6mm]
$R^+_\xi\equiv (Q^+)^{1/3}$&$Q^+_\xi\equiv f^+_{\rm gap}/f^+$\\ [1.6mm]
$Q^-_\xi\equiv f^-_{\rm gap}/f^-$&$Q^c_\xi\equiv f^c_{\rm gap}/f^c$\\ [1.6mm]
$Q_c \equiv B^2(f^+)^3/C^+$&$ Q_2\equiv (f^c/f^+)^{2-\eta} C^+/C_c$ \\ [1.6mm]
$P_m \equiv { T_p^\beta B /B_c}$ & 
$P_c \equiv - T_p^{2\beta\delta} C^+ /C_4^+$ \\ [1.6mm]
$R_p \equiv { C^+/C_p}$ & \\ [1.6mm]
\hline
\end{tabular}
\end{center}
\caption{
Amplitude-ratio definitions. 
}
\label{notationsur}
\end{table*}

In Table \ref{eqstdet} we report the universal amplitude ratios,
as derived by the approximate polynomial representations of the 
equation of state for $k=1,2,3,4$.
The reported errors are only due to the uncertainty of the
input parameters and do not include the systematic error of the
procedure, which may be determined by comparing the results of the
various approximations.  
In Table \ref{eqstdet} 
we also show results for $z_{\rm max}$, $x_{\rm max}$
and $y_{\rm max}$ which are the values of the
scaling variable $z$, $x$ and $y$ ($y$ was defined 
in Eq.~(\ref{Ey})) associated with the crossover line. 
As already mentioned in  Sec.~\ref{reseq},
we consider the $k=3$ results as our best estimates,
and report them in Table~\ref{summary}.

\begin{table}[tbp]
\footnotesize
\begin{center}
\caption{
Universal amplitude ratios obtained by taking different
approximations of the parametric function $h(\theta)$. 
}
\label{eqstdet}
\begin{tabular}{ccccc}
\hline\hline
\multicolumn{1}{c}{}&
\multicolumn{1}{c}{$k=1$}&
\multicolumn{1}{c}{$k=2$}&
\multicolumn{1}{c}{$k=3$}&
\multicolumn{1}{c}{$k=4$}\\
\hline \hline
$U_0$ & 0.5231(11) & 0.533(2) & 0.5319(25) &  0.529(6) \\

$U_2$ & 4.826(7) & 4.745(10) & 4.758(19)  & 4.78(5) \\  

$U_4$ & $-$9.73(3) & $-$8.85(6) & $-$9.0(2)  & $-$9.3(5) \\

$R_c^+$ & 0.05545(7) & 0.0570(1) & 0.0567(3)  & 0.0562(11) \\

$R_c^-$ & 0.021967(11) & 0.02253(3) & 0.02242(12) & 0.0222(4) \\

$R_4^+$ & 7.983(4) & 7.794(8) & 7.81(2) & 7.83(4) \\

$R_4^-$ & 92.15(13) & 94.13(13) & 93.6(6) & 92(2) \\

$R_\chi$ & 1.6779(11) & 1.658(2) & 1.660(4) & 1.665(10) \\

$U_2\,R_4^+$ & 38.52(5) & 36.98(10) & 37.1(2) & 37.4(6) \\

$R_4^+\,R_c^+$ & 0.4427(7) & 0.4444(7) & 0.443(2) & 0.440(6) \\

$P_m$ & 1.25203(6) & 1.2493(2) & 1.2498(6) & 1.251(2) \\

$P_c$ & 0.3831(3) & 0.3938(5) & 0.3933(7) & 0.3930(11) \\

$R_p$ & 1.9789(3) & 1.9658(6) & 1.9665(10) & 1.9671(16)  \\ 

$z_{\rm max}$  & 1.2443(4) & 1.2317(5) & 1.2322(8) & 1.2326(12) \\

$x_{\rm max}$  & 12.32(3) & 12.26(3) & 12.27(4) & 12.31(8) \\

$w_{\rm max}$ & 1.990(1) &  1.977(2) & 1.980(4) & 1.984(9) \\

$D(w_{\rm max})$ & 0.36179(4) &  0.36277(7) & 0.36268(14) & 0.3626(3) \\
\hline\hline
\end{tabular}
\end{center}
\end{table}

Estimates of universal ratios of amplitudes 
involving correlation-length amplitudes, such as 
$Q^+$, $R_\xi^+$, and $Q_c$, can be obtained using the
HT estimate of $g_4$. For instance
$Q^+ = R_4^+ R_c^+/g_4$.
Other universal ratios 
can be obtained by
supplementing the above results with the estimates of 
$w^2$ and  $Q^-_\xi$ (see Table \ref{notationsur}),
obtained by an analysis of the
corresponding low-temperature expansions \cite{CPRV-98,PV-98-g},
and the HT estimate of $Q^+_\xi$ (see Sec.~\ref{twopointf}).  
Moreover, using approximate
parametric representations of the correlation
length, see Refs.~\cite{CPRV-99,PV-review} for details,
one may also estimate 
the universal ratios $Q_\xi^c$ and $Q_2$ defined in
Table \ref{notationsur}.

\section{Low-momentum behavior of the structure factor}
\label{twopointf}

In this section we update the determination of the first few
coefficients that parametrize the low-momentum expansion of the 
scaling two-point function in the HT phase 
\cite{AF-73,CPRV-98,CPRV-99}
\begin{equation}
g(y) \equiv  {\chi\over \widetilde{G}(k)} =1 + y + \sum_{i=2}^\infty c_i y^i,
\end{equation}
where $y=k^2\xi^2$.

The coefficients $c_i$ 
can be related to the critical limit of appropriate dimensionless ratios of spherical
moments $m_{2j}$. See Ref.~\cite{CPRV-98} for details.
We have estimated the first few coefficients $c_i$ from the
corresponding series derived from the 25th-order expansions
of $m_{2j}$, using the analysis described in Sec.~\ref{IHTrj}.
The results for the three improved models and our final estimates 
are reported in Table~\ref{cires}.
Other interesting quantities are
\begin{eqnarray}
S_M&\equiv&M_{\rm gap}^2/M^2,\label{SMdef}\\
S_Z&\equiv& \chi M^2/Z_{\rm gap},\label{SZdef}
\end{eqnarray}
where $M_{\rm gap}$ (the mass gap of the theory) and $Z_{\rm gap}$ 
determine the long-distance behavior of the two-point function: 
\begin{equation}
G(x)\approx  {Z_{\rm gap}\over 4\pi |x|} e^{-M_{\rm gap}|x|}.
\label{largexbehavior}
\end{equation}
As discussed in Refs.~\cite{CPRV-98,CPRV-99}, one may estimate $S_M$ and
$S_Z$ from $c_2$, $c_3$, and $c_4$.  Indeed, we have
\begin{eqnarray}
S_M &=& 1 + c_2 - c_3  + c_4 + 2 c_2^2 + ... \label{SMest}\\
S_Z &=& 1 - 2 c_2 + 3 c_3 - 4 c_4 - 2 c_2^2 + ... \label{SZest}
\end{eqnarray}
where the ellipses indicate contributions that are negligible with
respect to $c_4$. Therefore, one finds $S_M=0.999601(6)$ and
$S_Z=1.000810(13)$.
From the result for $S_M$, one obtains $Q^+_\xi\equiv f^+_{\rm gap}/f^+=1.000200(3)$.

A more detailed analysis of the behavior of the structure factor for all 
momenta can be found in Ref.~\cite{MPV-02}.

\begin{table}[tbp]
\footnotesize
\begin{center}
\caption{
Estimates of the coefficients $c_i$, $i=2,3,4$,  of the low-momentum
expansion of the structure factor.
}
\label{cires}
\begin{tabular}{cllll}
\hline\hline
\multicolumn{1}{l}{}&
\multicolumn{1}{c}{$\phi^4$}&
\multicolumn{1}{c}{$\phi^6$}&
\multicolumn{1}{c}{spin-1}&
\multicolumn{1}{c}{final estimates}\\
\hline \hline
$c_2$&$-0.390(7)\times 10^{-3}$ & $-0.390(6)\times 10^{-3}$ &
$-0.389(12)\times 10^{-3}$&  $-0.390(6)\times 10^{-3}$ \\
 
$c_3$& $0.882(8)\times 10^{-5}$ &   $0.882(6)\times 10^{-5}$ &
$0.88(4)\times 10^{-5}$ &   $0.882(6)\times 10^{-5}$ \\   

$c_4$& $-0.4(1)\times 10^{-6}$ &  $-0.4(1)\times 10^{-6}$ &
$-0.4(1)\times 10^{-6}$ &   $-0.4(1)\times 10^{-6}$ \\
\hline\hline
\end{tabular}
\end{center}
\end{table}


\newpage
\begin{thebibliography}{199}

\bibitem{PV-review}
A. Pelissetto and E. Vicari,
``Critical Phenomena and Renormalization-Group Theory'',
e-print cond-mat/0012164.

\bibitem{Guttmann-rev-89}
A.~J.~Guttmann, in 
{\em Phase Transitions and Critical Phenomena}, Vol.\ 13, 
edited by C.~Domb and J.~Lebowitz
(Academic, New York, 1989).

\bibitem{ZJ-79}
J. Zinn-Justin, J. Phys. (France) {\bf 40}, 969 (1979);
{\bf 42}, 783 (1981).

\bibitem{Nickel-82}
B.~G.~Nickel, in {\em Phase Transitions},
M.~L\'evy, J.~C.~Le~Guillou, and J.~Zinn-Justin eds.,
(Plenum, New York and London, 1982).

\bibitem{Gaunt-82}
D.~S.~Gaunt, in {\em Phase Transitions},
M.~L\'evy, J.~C.~Le~Guillou, and J.~Zinn-Justin eds.,
(Plenum, New York and London, 1982).

\bibitem{CFN-82} 
J.-H.~Chen, M.~E.~Fisher, and B.~G.~Nickel,
Phys.\ Rev.\ Lett.\ {\bf 48}, 630  (1982).

\bibitem{Adler-83} 
J.~Adler, J.\ Phys.\ A {\bf 16}, 3585 (1983).

\bibitem{GR-84} M.~J.~George and J.~J.~Rehr,
Phys.\ Rev.\ Lett.\ {\bf 53}, 2063 (1984).

\bibitem{FC-85} 
M.~E.~Fisher and J.~H.~Chen,
J.\ Physique {\bf 46}, 1645 (1985).

\bibitem{BC-00} 
P.~Butera and M.~Comi, 
Phys. Rev. B {\bf 62}, 14837 (2000).

\bibitem{Roskies-81} 
R.~Z.~Roskies,
Phys.\ Rev.\ B {\bf 24}, 5305 (1981).

\bibitem{AMP-82} 
J.~Adler, M.~Moshe, and V.~Privman,
Phys.\ Rev.\ B {\bf 26}, 1411 (1982); 
B {\bf 26}, 3958 (1982).

\bibitem{BC-97} 
P.~Butera and M.~Comi, 
Phys.\ Rev.\ B {\bf 56}, 8212 (1997).

\bibitem{PV-98-g} 
A.~Pelissetto and E.~Vicari,
Nucl.\ Phys.\ B {\bf 519}, 626 (1998).

\bibitem{BC-98}
 P.~Butera and M.~Comi, 
Phys.\ Rev.\ B {\bf 58}, 11552 (1998).

\bibitem{MJHMJG-00}
D. MacDonald, S. Joseph, D. L. Hunter,
L. L. Moseley, N. Jan, and A. J. Guttmann,
J. Phys. A {\bf 33}, 5973 (2000).

\bibitem{BC-02}
 P.~Butera and M.~Comi, 
``Critical universality and hyperscaling revisited
for Ising models of general spin using extended
high-temperature series'', e-print hep-lat/0112049.

\bibitem{NR-90}
B.~G.~Nickel and J.~J.~Rehr,
J.\ Stat.\ Phys.\ {\bf 61}, 1 (1990).

\bibitem{CPRV-99}
M.~Campostrini, A.~Pelissetto, P.~Rossi, and E.~Vicari,
Phys.\ Rev.\ E {\bf 60}, 3526 (1999).

\bibitem{BFMM-98}
H.~G.~Ballesteros, L.~A.~Fern\'andez, V.~Mart\'{\i}n-Mayor,
and A.~Mu\~noz Sudupe, Phys.\ Lett.\ B {\bf 441}, 330 (1998).

\bibitem{HPV-99}
M.~Hasenbusch, K.~Pinn, and S.~Vinti,
Phys.\ Rev.\ B {\bf 59}, 11471 (1999). 

\bibitem{BFMMPR-99}
H.~G.~Ballesteros, L.~A.~Fern\'andez, V.~Mart\'{\i}n-Mayor,
A.~Mu\~noz Sudupe, G.~Parisi, and J.~J.~Ruiz-Lorenzo,
J.~Phys.\ A {\bf 32}, 1  (1999).

\bibitem{Hasenbusch-99}
M.~Hasenbusch, J.~Phys.\ A {\bf 32}, 4851 (1999).

\bibitem{HT-99}
M.~Hasenbusch and T.~T\"or\"ok,  J.~Phys.\ A {\bf 32}, 6361 (1999).

\bibitem{CHPRV-01}
M.~Campostrini, M. Hasenbusch, A.~Pelissetto, P.~Rossi, and E.~Vicari,
Phys. Rev. B {\bf 63}, 214503 (2001).

\bibitem{Hasenbusch-01}
M. Hasenbusch,
J. Phys. A {\bf 34}, 8221 (2001).

\bibitem{CHPRV-02}
M.~Campostrini, M. Hasenbusch, A.~Pelissetto, P.~Rossi, and E.~Vicari,
cond-mat/0110336.

\bibitem{CPRV-00}
M.~Campostrini, A.~Pelissetto, P.~Rossi, and E.~Vicari,
Phys.\ Rev.\ B {\bf 61}, 5905 (2000).

\bibitem{CPRV-00-2}
M.~Campostrini, A.~Pelissetto, P.~Rossi, and E.~Vicari,
Phys.\ Rev.\ B {\bf 62}, 5843 (2000).

\bibitem{bcc-series}
There are also quite long series on the body-centered cubic lattice:
for $\chi$ and $m_2$, 21 orders were computed for the Klauder,
double-Gaussian, and Blume-Capel models for generic values of the
coupling in Ref.~\cite{NR-90}; 25 terms were computed for a generic
model in Ref.~\cite{Campostrini-01}.

\bibitem{phi4est}
This is the estimate used in Ref.~\cite{CPRV-99}, which was 
derived from the MC results  of
Ref.~\cite{Hasenbusch-99}. There, the result
$\lambda^*=1.095(12)$ was obtained by fitting the data for the lattices 
of size $ L \ge 16$.
Since fits using also data for smaller lattices, i.e.,
with  $L \ge 12$ and  $L \ge 14$, gave consistent results,
one might expect that the systematic error is at most as large as 
the statistical one \cite{Hasenbusch-pc}.

\bibitem{Hasenbusch-99-h}
M.~Hasenbusch, Habilitationsschrift, Humboldt-Universit\"at zu
Berlin, 1999; Int. J. Mod. Phys. C {\bf 12}, 911 (2001).

\bibitem{foot-omegaNR}
The exponent $\omega_{\rm NR}$ is associated with the leading 
nonrotationally invariant scaling corrections: 
see Ref. \cite{CPRV-98} for a precise definition.

\bibitem{CPRV-98} 
M.~Campostrini, A.~Pelissetto, P.~Rossi, and E.~Vicari,
Phys.\ Rev.\ E {\bf 57}, 184 (1998).



\bibitem{BB-85}
C.~Bagnuls and C.~Bervillier,
Phys.\ Rev.\ B {\bf 32}, 7209 (1985);
``Classical-to-critical crossovers from field theory'',
e-print hep-th/0112209.

\bibitem{BBMN-87}
C.~Bagnuls, C.~Bervillier,
D.~I.~Meiron, and B.~G.~Nickel,
Phys.\ Rev.\ B {\bf 35}, 3585 (1987);
addendum-erratum e-print hep-th/0006187 (2000).


\bibitem{LF-89}
A.~J.~Liu and M.~E.~Fisher,
Physica A {\bf 156}, 35 (1989).

\bibitem{GKM-96}
C.~Gutsfeld, J.~K\"uster, and G.~M\"unster,
Nucl.\ Phys.\ B {\bf 479}, 654 (1996).

\bibitem{ZF-96}
S.-Y.~Zinn and M.~E.~Fisher,
Physica A {\bf 226}, 168 (1996).

\bibitem{CH-97}
M.~Caselle and M.~Hasenbusch,
J.\ Phys.\ A {\bf 30}, 4963 (1997).

\bibitem{GZ-97}
R.~Guida and J.~Zinn-Justin,
Nucl.\ Phys.\ B {\bf  489}, 626 (1997).

\bibitem{FZ-98}
M.~E.~Fisher and S.-Y.~Zinn, J.\ Phys.\ A {\bf 31}, L629 (1998).

\bibitem{GZ-98} R.~Guida and J.~Zinn-Justin,
J.~Phys. A {\bf 31}, 8103 (1998).

\bibitem{HP-98}
M.~Hasenbusch and K.~Pinn,
J.\ Phys.\ A {\bf 31}, 6157 (1998).

\bibitem{LMSD-98}
S.~A.~Larin, M.~M\"onnigman, M.~Str\"osser, and V.~Dohm,
Phys.\ Rev.\ B {\bf 58}, 3394 (1998).

\bibitem{PV-98}
A.~Pelissetto and E.~Vicari, Nucl.\ Phys.\ B {\bf 522}, 605 (1998);
B {\bf 575}, 579 (2000).


\bibitem{BST-99}
H. W. J. Bl\"ote, L. N. Shchur and A. L. Talapov,
Int. J. Mod. Phys. C {\bf 10}, 137 (1999).

\bibitem{ZJ-01}
J. Zinn-Justin,
Phys. Rep. {\bf 344}, 159 (2001).

\bibitem{PV-99}
A.~Pelissetto and E.~Vicari, Nucl.\ Phys.\ B {\bf 540}, 639 (1999).

\bibitem{NGMJ-01}
A. W. Nowicki, Madhujit Ghosh, S. M. McClellan,
and D. T. Jacobs,
J. Chem. Phys. {\bf 114}, 4625 (2001).

\bibitem{MISTE}
M. Barmatz, ``MISTE Science Requirements Document," 
Tech. Rep. JPL D-17083, JPL (1999).

M. Barmatz, I. Hahn, and F. Zhong,
``Progress in the Development of the MISTE Flight Experiment,"
to appear in Proceedings of 2000 NASA/JPL Workshop on 
Fundamental Physics in Microgravity, edited by D. Strayer.



\bibitem{Wortis-74} 
M.~Wortis, ``Linked cluster expansion'', in {\em Phase Transitions and
Critical Phenomena}, vol.~3, edited by C.~Domb and M.~S.~Green
(Academic Press, London, 1974).

\bibitem{Campostrini-01}
M. Campostrini, J. Stat. Phys. {\bf 103}, 369 (2001).

\bibitem{Fisher-62}
M.~E.~Fisher, Philos.\ Mag.\ {\bf 7}, 1731 (1962).

\bibitem{CPRM}
In the 
so-called critical-point renormalization method  
(see Ref.\ \cite{int-appr-ref} and references therein), given two
series $D(x)$ and $E(x)$ that are singular at the same point $x_0$,
$D(x)=\sum_i d_i x^i\sim (x_0-x)^{-\delta}$ and 
$E(x)=\sum_i e_i x^i\sim (x_0-x)^{-\epsilon}$, one constructs a new
series $F(x)=\sum_i (d_i/e_i)x^i$.  The function $F(x)$ is singular at
$x=1$ and for $x\to 1$ behaves like $F(x)\sim (1-x)^{-\phi}$, where
$\phi = 1+ \delta - \epsilon$.  Therefore, the difference
$\delta-\epsilon$ can be obtained by analyzing the expansion of $F(x)$
by means of biased approximants with a singularity at $x_c=1$. 

\bibitem{NR-84}
K. E. Newman and E. K. Riedel, 
Phys. Rev. B {\bf 30}, 6615 (1984).

\bibitem{foot-correction}
Using the results of Sec. \ref{matching}, we can estimate the
size of a correction $n^{-2-\Delta}$. Since $A_0\approx 0.5246$,
$|A_{1+\Delta}|\lesssim 10^{-2}$, we have 
$\beta_c^{(n)}\approx \beta_c (1 + a n^{-2-\Delta})$ with 
$|a|\lesssim 2\times 10^{-2}$. Thus, the correction may be 
small and give a negligible contribution at present values of $n$.



\bibitem{singatcoexcurve}
M.~E.~Fisher, Physics {\bf 3}, 255 (1967);

A.~F.~Andreev, Sov.\ Phys.\ JETP {\bf 18}, 1415 (1964);

M.~E.~Fisher and B.~U.~Felderhof, Ann.\ Phys.\ 
(NY) {\bf 58} 176, 217 (1970);

S.~N.~Isakov, Comm.\ Math.\ Phys.\ {\bf 95}, 427 (1984).

\bibitem{foot-series}
Using the available series, we obtain
$g_4=\beta^{-3/2}\sum_{i=0}^{n_{\rm max}} c_i \beta^i$ and
$r_{2j} = \sum_{i=0}^{n_{\rm max}} c_i \beta^i$,
with $n_{\rm max} = 20,19,17$ for $g_4$, $r_6$, and $r_8$.

\bibitem{int-appr-ref}
D.~L.~Hunter and G.~A.~Baker, Jr.,
Phys.\ Rev.\ B {\bf 7}, 3346 (1973); B {\bf 7}, 3377 (1973); 
B {\bf 19}, 3808 (1979);

M.~E.~Fisher and H.~Au-Yang,
J.\ Phys.\ A {\bf 12}, 1677 (1979); Erratum A {\bf 13}, 1517 (1980);

A.~J.~Guttmann and G.~S.~Joyce, J.\ Phys.\ A {\bf 5}, L81 (1972);

J.~J.~Rehr, A.~J.~Guttmann, and G.~S~Joyce, 
J.\ Phys.\ A {\bf 13}, 1587 (1980).

\bibitem{CPRVan}
In Ref.~\cite{CPRV-99} 
we used Pad\'e, Dlog-Pad\'e, and IA1, requiring that they were not
singular at $\beta_c$. We obtained $g_4=23.49(4)$. 

\bibitem{CHPV-01}
M. Caselle, M. Hasenbusch, A. Pelissetto, and E. Vicari,
J. Phys. A {\bf 34}, 2923  (2001).

\bibitem{Schofield-69}
P. Schofield, Phys. Rev. Lett. {\bf 22} (1969) 606.

\bibitem{SLH-69}
P. Schofield, J. D. Lister, and J. T. Ho,
Phys. Rev. Lett. {\bf 23}  (1969) 1098.

\bibitem{Josephson-69}
B. D. Josephson, J. Phys. C: Solid State Phys. {\bf 2} (1969) 1113.


\bibitem{otherparam}
H. B. Tarko and M. E. Fisher, Phys. Rev. B 11, 1217 (1975);
   
M. E. Fisher, S.-Y. Zinn, and P. J. Upton, Phys. Rev. B 59, 14533 (1999);

M. A. Anisimov and J. V. Sengers, in {\em Equations of State for 
Fluids and Fluid Mixtures}, edited by 
J. V. Sengers, R. F. Kayser, C. J. Peters, and H. J. White, Jr.
(Elsevier, Amsterdam, 2000).

\bibitem{KR-67}
J. S. Kouvel and D. S. Rodbell,
Phys. Rev. Lett. {\bf 18}, 215 (1967).

\bibitem{BK-97}
P. D. Babu and S. N. Kaul,
J. Phys.: Condens. Matter {\bf 9}, 7189 (1997).

\bibitem{AN-67}
A. Arrott and J. E. Noakes,
Phys. Rev. Lett. {\bf 19}, 786 (1967).

\bibitem{AF-73}
M. E. Fisher and A. Aharony,
Phys. Rev. B {\bf 10}, 2818 (1974).

\bibitem{MPV-02}
V.~Mart\'{\i}n-Mayor, A. Pelissetto, and E. Vicari, 
``Critical structure factor in Ising systems,''
in preparation.

\bibitem{Hasenbusch-pc}
M.~Hasenbusch, private communication.

\end{thebibliography}
\end{document}